\documentclass[final,5p,twocolumn,authoryear]{elsarticle}

\usepackage{lineno}

\usepackage{multicol}
\usepackage{url,multirow,rotating,color}
\usepackage[table]{xcolor}

\usepackage{listings}
\usepackage{subcaption}
\usepackage[ruled]{algorithm2e}

\usepackage{bm}
\usepackage{tikz}

\newcommand{\ie}{\textit{i.e.,}\xspace}
\newcommand{\eg}{\textit{e.g.,}\xspace}

\newcommand{\new}[1]{\textcolor{black}{#1}}
\def\ourtechnique{\textsc{GEML}\xspace}

\journal{Journal of Systems and Software}
\date{January 24, 2021}

\usepackage{fancyhdr}
\begin{document}

\begin{frontmatter}

\title{\ourtechnique: A Grammar-based Evolutionary Machine Learning Approach for Design-Pattern Detection}



\author[uco]{Rafael Barbudo\corref{x}}\ead{rbarbudo@uco.es}
\author[uco]{Aurora Ram\'irez\corref{x}}\ead{aramirez@uco.es}
\author[vt]{Francisco Servant\corref{x}}\ead{fservant@vt.edu}
\author[uco]{Jos\'e Ra\'ul Romero\corref{corauthor}}\ead{jrromero@uco.es}
\cortext[corauthor]{Corresponding author. Tel. +34 957 21 26 60}

\address[uco]{Department of Computer Science and Numerical Analysis, University of C\'ordoba, 14071, C\'ordoba, Spain}
\address[vt]{Department of Computer Science, Virginia Tech, VA 24060, Blacksburg, USA}

\begin{abstract}
Design patterns (DPs) are recognised as a good practice in software development. However, the lack of appropriate documentation often hampers traceability, and their benefits are blurred among thousands of lines of code. Automatic methods for DP detection have become relevant but are usually based on the rigid analysis of either software metrics or specific properties of the source code.  We propose \ourtechnique, a novel detection approach based on evolutionary machine learning using software properties of diverse nature. Firstly, \ourtechnique makes use of an evolutionary algorithm to extract those characteristics that better describe the DP, formulated in terms of human-readable rules, whose syntax is conformant with a context-free grammar. Secondly, a rule-based classifier is built to predict whether new code contains a hidden DP implementation. \ourtechnique has been validated over five DPs taken from a public repository recurrently adopted by machine learning studies. Then, we increase this number up to 15 diverse DPs, showing its effectiveness and robustness in terms of detection capability. An initial parameter study served to tune a parameter setup whose performance guarantees the general applicability of this approach without the need to adjust complex parameters to a specific pattern. Finally, a demonstration tool is also provided.
\end{abstract}


\end{frontmatter}


\section{Introduction}
\label{sec:intro}

Design patterns (DPs) are reusable template solutions that address recurrent software-design problems. The adoption of DPs is a best practice for programmers with the goal of improving the quality of software products --- in terms of their maintainability, elegance, flexibility and understandability~\citep{gamma1995}. Given such benefits, and considering that manual inspection is an error-prone and time-consuming process, automatic Design Pattern Detection (DPD) has become a prominent area in the reverse-engineering research field~\citep{mayvan2017}. By automatically identifying the adoption of DPs, DPD techniques can improve the understanding of the design decisions in software systems, as well as the processes of redocumenting them, reimplementing them, and reusing them.

Different techniques have been proposed in the research literature to automate DPD, most of which search for particular structures in static code~\citep{bafandeh2017}. In this case, the code structures defining DPs need to be predetermined by experts into a knowledge-base, as they usually are specific to the codebase of study. Having the expert defined these code structures may impose rigidity to the detection technique, and design patterns should be reinterpreted for particular contexts.

To reduce this limitation, machine learning (ML) techniques were proposed for DPD, \eg \cite{ferenc2005}. These techniques are more easily adapted to different codebases because they learn from a collection of representative examples --- and thus can recognise diverse implementations by simply replacing or extending such analysed collection of examples. In particular, ML-based approaches provide mechanisms to learn the structural and behavioural properties of the source code, as well as software metrics, that best describe the DP. Even so, despite the fact that DPs can be described considering these multiple aspects, current ML approaches consider either software metrics or code properties --- structural, behavioural or both.

Unfortunately, the detection performance of existing ML-based approaches for DPD is affected by the design pattern under analysis, often requiring its specific parameter configuration. This characteristic makes them harder to adopt in practice: tuning parameters requires considerable effort by software engineers, since machine learning is often not their main area of expertise. Furthermore, the results provided by certain ML-based techniques like neural networks are difficult to understand and interpret by the human expert~\citep{Mori19}, thus making their recommendations less likely to be trusted~\citep{dzindolet2003role}.

In this paper, we present \ourtechnique, a novel machine-learning-based approach for DPD from static code. More specifically, \ourtechnique is founded on associative classification (AC), a ML technique for which we have implemented a solution based on grammar-guided genetic programming (G3P). Our goal is to achieve the benefits of ML-based proposals while also addressing their limitations by using G3P as a basis of our approach. Like other ML-based proposals, \ourtechnique learns from DP examples, giving it the ability to capture diverse DP implementations. As for the limitations of previous ML proposals, \ourtechnique has been designed to promote extensibility, readability and flexibility.

For extensibility, \ourtechnique includes a customisable collection of design microstructures whose potential to identify DP instances is automatically determined by the evolutionary algorithm during the learning phase. For readability, \ourtechnique builds a rule-based classifier guided by a configurable context-free grammar that declares the syntax of the rules. Rules, as a well-established mechanism to encode human knowledge~\citep{Grosan11}, describe the distinctive characteristics of DP instances in a more comprehensible way than the outcomes produced by black-box models~\citep{Kotsiantis06}. And for flexibility, \ourtechnique is applicable to each DP without requiring different algorithms or parameter configurations, since it searches for optimal rules to describe each particular DP.

In addition to these benefits, in our experiments, \ourtechnique also provided accuracy levels that outperformed other available techniques, providing competitive results even when very few samples are provided. Furthermore, \ourtechnique is able to maintain a stable behaviour with just a single general configuration, which we believe will make it more beneficial for software engineers in practical settings.

In short, \ourtechnique allows developers to execute it out-of-the-box without having to configure it for each DP, and customise its elements and parameters to gain detection power in particular scenarios. However, these desirable qualities would only make \ourtechnique beneficial in practice if it also provided (reasonably) similar performance as other existing techniques. Thus, we focus our evaluation on studying the effectiveness provided by \ourtechnique.

We perform several experiments in our evaluation. Firstly, we study \ourtechnique's detection performance in depth through a sensitivity analysis. Although these kinds of analysis are laborious, they become essential to comprehend the internals of artificial intelligence techniques like those applied here. In particular, we aim to understand the extent to which each algorithmic component and its configuration contributed to detecting each individual DP. In this way we are able to determine the best detection capability offered by \ourtechnique. We also study the selection options and effectiveness of \ourtechnique if its configuration was customised in terms of the design microstructures eligible by the software engineer and software metrics --- called operators ---  that describe the properties that best characterise each DP. Then, we also experimented with the alternative scenario in which software engineers did have the knowledge and time to customise it separately for each DP. As additional benefits of customising \ourtechnique individually for each DP, we expect that practitioners would also observe lower runtimes. Finally, we measure the detection performance of the general configuration of \ourtechnique. We anticipate that this would be the most common (and simpler) usage scenario for practitioners. 

We also analyse how \ourtechnique behaves when the general configuration is set but the training conditions change. In this case, we consider up to 15 DPs --- the highest number of DPs studied from a ML perspective --- at the cost of reducing the number of available training samples. Even so, our results show that \ourtechnique is able to infer detection rules in all cases, not requiring any adaptation regarding its internal components and configuration. We also compare \ourtechnique's detection performance to other DPD methods, including both ML and non-ML techniques. \ourtechnique provided higher accuracy and $F_1$ score than MARPLE~\citep{zanoni2015} --- a well-known ML-based approach --- for four out of the five DPs available for comparison (with up to 35\% improvement in $F_1$ for one of them). \new{Against other non-ML-based techniques, \ourtechnique also recover more DP instances when validating with JHotDraw, a frequently studied project in DPD literature.} Lastly, we analyse the strengths and weaknesses of \ourtechnique with respect to SSA and Ptidej, the two reference DPD tools most frequently used for comparative purposes. In this sense, \ourtechnique is highly competitive since it correctly identifies a higher number of DP implementations and support DPs whose detection is not available in these tools. \ourtechnique also overcomes some limitations of these tools, such as the absence of the classes implementing some roles of the DP and the excessive number of false positives. In short, the results of our evaluation show that \ourtechnique is a practical approach for DPD: it improves the effectiveness of current methods, while returning readable outcomes. Furthermore, the possibility of choosing the code properties more relevant for learning brings flexibility to the DPD process.

This paper provides the following contributions:
\begin{itemize}
	\item a novel DPD technique (\ourtechnique) based on machine learning with grammar-guided genetic programming that is able to provide a single configuration for detecting 15 DPs and returning human-readable rules;
	\item the provision of a collection of design microstructures and metrics, in terms of operators that are derived by the context-free grammar (CFG) guiding the G3P algorithm and defined with both categorical and numerical values;
    \item an analysis of which design microstructures and software metrics best represent 
    each DP.
    \item a experimental evaluation of \ourtechnique under different training and configuration scenarios, finding highly competitive results when compared against other ML and non-ML approaches;
    \item a research demonstration tool to support software engineers in executing \ourtechnique and customising it for individual DPs.
\end{itemize}

The rest of the paper is organised as follows. Main concepts and terminology related to the applied techniques are introduced in Section~\ref{sec:background}. Section~\ref{sec:related} presents the related work, and Section~\ref{sec:approach} provides a detailed description of our DPD model. Section~\ref{sec:settings} details the experimental methodology and framework. Then, the \new{three} experiments conducted are discussed in Section~\ref{sec:experiment1},~\ref{sec:experiment2} and~\ref{sec:experiment3}. Section~\ref{sec:tool} describes the demonstration tool provided as additional material supporting this approach. Finally, threats to validity and concluding remarks are presented in Sections~\ref{sec:threads} and~\ref{sec:concluding}, respectively.


\section{Theoretical Background}\label{sec:background}

Design patterns are descriptions of communicating objects and classes that are customised to solve a general design problem in a particular context~\citep{gamma1995}. They differ in the number and purpose of their defining roles, each one describing a specific task to be performed. Consequently, a DPD method does not only identify the code elements implementing the design pattern, but also the roles they play within the pattern structure. Notice that the definition of these elements and how they relate to each other might depend on the particularities of the programming language, \eg Java allows implementing explicit interfaces but C++ does not. Besides, a given role could be played by more than one code element. Therefore, since design patterns are general and adaptable solutions, the presence of certain properties or a predefined structure for roles cannot be assumed. Their diverse nature, \ie behavioural, creational or structural, suggests that different properties might be needed to characterise a given pattern, what definitely may hamper the detection process. 

This section explains the most relevant theoretical concepts related to the techniques required by \ourtechnique to conduct the detection procedure. This is built on the basis of associative classification, \ie a ML-based approach founded on the use of \textit{if-then} rules over a classification approach looking for an understandable detection model~\citep{thabtah2007}. Internally, the classifier is constructed using an evolutionary technique, G3P, specially conceived to evolve computer programs according to a grammar, whose derivations and constraints define how these programs --- \ie a potential rule of the \ourtechnique's detection model --- are built.

\subsection{Associative Classification}\label{sec:background_ac}

Machine learning provides computational methods that allow learning from vast collections of data. ML techniques can be mostly divided into two major groups, unsupervised and supervised. Unsupervised techniques explore the data samples to find interesting and meaningful patterns describing them. A well-known technique within unsupervised learning is association rule mining (ARM)~\citep{agrawal1993}, where patterns are represented as a set of association rules. Formally, let I = \{$i_1$, ..., $i_n$\} be a set of items, and let A and C be itemsets, \ie A = \{$i_1$, ..., $i_j$\} and C = \{$i_1$, ..., $i_k$\}, an association rule is an implication of the type $A \rightarrow C$ where A $\subset$ I, C $\subset$ I, and A $\cap$ C = $\emptyset$.

In ARM, quality measures are computed to determine the quality of the produced rules, the support and confidence being the most representative measures. On the one hand, the support (Eq.~\ref{eq:support}) indicates how frequently a rule is satisfied within a given set of data samples ($D$). On the other hand, the confidence (Eq.~\ref{eq:confidence}) measures the proportion of samples that satisfy the consequent from those already satisfying the antecedent. In both equations, $s$ represents each data sample within $D$.

\begin{equation}
supp(A \rightarrow C) = \frac{|\left\lbrace A \cup C \subseteq s \right\rbrace|}{|D|} \;\;s \in D
\label{eq:support}
\end{equation}

\begin{equation}
conf(A \rightarrow C) = \frac{|\left\lbrace A \cup C \subseteq s\right\rbrace|}{|\left\lbrace A\subseteq s\right\rbrace|} \;\;s \in D
\label{eq:confidence}
\end{equation}

Unlike unsupervised learning, supervised techniques make predictions based on the information extracted from past data. The purpose of a classification algorithm is to assign a predefined category, referred as class, to an unknown data sample. In particular, associative classification makes use of ARM techniques to build rule-based classifiers~\citep{thabtah2007}. With this aim, class association rules (CARs)~\citep{liu1998}, those in which the consequent determines the class, are generated. In general, AC can be viewed as a two-step process: ($1$) an ARM-based method is applied to mine a set of CARs; and ($2$) the set of CARs is pruned to exclusively select those rules that will be definitely constitute the classification model.

\subsection{Grammar-guided Genetic Programming (G3P)}\label{sec:background_g3p}

Inspired by the principles of natural evolution, evolutionary computation creates a population of individuals, each one representing a potential solution to the problem, which are then ``evolved'' during a number of generations~\citep{eiben2015}. An individual is characterised by its genotype, \ie the computational structure used to encode the solution, and its phenotype, \ie its real-world representation. In addition, a domain-specific fitness function has to be defined to assess the quality of an individual. In a general schema, for each generation, individuals are selected and modified with the aim of producing better individuals than their predecessors. More specifically, the algorithm selects a subset of the population to act as parents, often based on their fitness value. Parents are then combined to generate offspring by means of a crossover operator. Mutation can also be applied to the obtained individuals, looking for more diverse solutions. At the end of the generation, the best individuals survive and create the next population.

Genetic programming (GP) is a special type of EC technique in which individuals are encoded using tree structures~\citep{koza1992}. Originally conceived to evolve computer programs, GP requires specialised operators to manipulate tree-based solutions of different shapes and sizes. In particular, G3P is an extension of GP in which a CFG describes the syntactic constrains that must be satisfied by any valid individual. A CFG is defined by a four-tuple $\lbrace S,\sum_N, \sum_T, P\rbrace$, where $S$ is the root symbol, $\sum_N$ is the set of non-terminal symbols, $\sum_T$, is the set of terminal symbols and $P$ is the set of production rules. A production rule indicates how a non-terminal symbol can be rewritten into one of their derivations until the expression only contains terminal symbols. Formally, it can be expressed as $a \rightarrow B$, where $a \in \sum_N$ and $B \in \lbrace\sum_N \cup \sum_T\rbrace^*$.

In G3P, each individual is created by deriving a different sequence of production rules, represented as a derivation tree. The elements of the CFG are also considered during the application of crossover and mutation to guarantee the production of valid solutions. G3P has been used to mine association rules~\citep{luna2012}, where the grammar formally defines the structure of the rule in terms of itemsets. In this context, each individual represents an association rule, and the evolutionary process is oriented towards finding a set of high-quality rules according to support and confidence criteria.

\section{Related Work}\label{sec:related}

Multiple techniques have been proposed for automatic detection of structural, behavioural and creational design patterns, using both static and dynamic analysis of source code. Also, both types of techniques can be combined together depending on the pattern to be detected, \eg by inspecting both class definitions and object collaborations~\citep{Ng2010,DeLucia2018}. Given that \ourtechnique performs static analysis, we mostly focused on these approaches for DP detection, with special attention to those proposals based on machine learning. In addition, even though the majority of studies take source code as input, other authors have explored detection at design stages too, \eg using UML class diagrams~\citep{DiMartino2016}.

Within the scope of reasoning techniques, the Pat system~\citep{kramer1996} is considered a pioneering work based on declarative logic programming. It defines the structural properties of DPs and the software project as Prolog facts. Then, the Prolog search engine is used to look for exact matches of these facts, which are identified as new DP instances. Similarly, the use of Prolog facts together with the use of meta-patterns~\citep{pree1997}, which are common structures of DPs, were used by~\cite{hayashi2008}. In this context, the FINDER tool~\citep{dabain2015} integrates facts related to class structure and method invocations. Then, it filters out the fact base using predefined detection scripts. Fuzzy logic has been also used to deal with incomplete knowledge~\citep{niere2002} in a attempt to make the detection process more flexible. In this approach, a fuzzy weight is assigned to each either structural or behavioural property thus reporting a level of confidence to each potential match. However, fuzzy-based approaches require an extensive base of knowledge to represent the diverse implementations. Given that such a base of knowledge has to be constructed by a group of experts, outcomes from the detection process could be biased.

Alternatively, techniques based on similarity scoring adopt graphs to represent structural information. The SSA tool~\citep{tsantalis2006} searches for substructures that correspond to a predefined template of the graph describing the structure of a certain DP, \ie the explicit representation of the relationships between roles. Mechanisms to deal with approximate matches can be applied too, such as considering a small variation range in the value of some properties. Similarly, behavioural properties like message passing are often considered to reduce false positives~\citep{dong2009}. In a different approach~\citep{yu2015}, a set of substructures are searched prior to the identification of the DP implementations. Then, method signatures are compared against a set of predefined templates describing the DPs to refine the results. In this vein, two other state-of-the-art approaches focus on improving the prediction performance of the substructures search process~\citep{mayvan2017,yu2018}. While the former reduces the search space by conveniently partitioning the project graph, the latter defines ordered sequences to guide the search in such an order that the most representative classes of the pattern are discovered first, filtering irrelevant classes at an early stage. DPF is a model-based graph matching tool for which a meta-model and a domain-specific language are proposed to specify the structural and behavioural relationships describing DPs~\citep{Bernardi2014}. Another popular tool, PINOT~\citep{Shi2006}, uses graphs as an abstract representation of methods taking part in candidate DP implementations, from which control flow is statically analysed.

Formal methods have been explored in the context of DPD too. More specifically, formal concept analysis has been applied to find groups of classes sharing common structural relations that represent candidate design patterns~\citep{tonella2001}. This approach was later extended introducing some additional contributions like a filtering phase of candidate patterns and proposing a language-independent variant~\citep{arevalo2004}. A case study was also conducted to detect structural patterns within two subsystems of a printer controller~\citep{wierda2009}. However, formal approaches can be computationally expensive due to the exponential growth of formal concepts~\citep{poelmans2013}. 

Design patterns have also been expressed in terms of features and documented using annotations, from which automatic analysers can be built~\citep{Rasool2010, Rasool2011}. As the authors acknowledge, incomplete definitions or inappropriate semantics negatively impact the detection process. Relations defined on the basis of a visual language represent another way to specify the properties of structural DPs~\citep{DeLucia2009}. In order to reduce false positives of the detection process, this method supports the definition of negative criteria, \ie those properties that do not indicate the presence of a DP. Then it performs a low-level analysis of the relationships between classes in a second step.

Other authors have formulated DPD as a constraint satisfaction problem, defining those structural and behavioural conditions that each particular DP should satisfy. This type of method requires the manual definition and formalisation of the relationships between roles. DeMIMA~\citep{Gueheneuc2008} and the DPJF tool~\citep{Binun2012} are representative examples of this approach, both supporting constraint relaxation. The former, which provides explanations of satisfied and non-satisfied constraints, has been later extended to include numerical properties as a way to reduce the number of candidates per role~\citep{Gueheneuc2010}. DeMIMA and its extensions are available within the Ptidej tool suite.

ML has been applied to support the DPD process, mainly building classification models to characterise DPs through the inspection of code implementations. These proposals differ in terms of the role that ML plays in the detection process, the properties of the source code used for learning, and the specific algorithm applied. A first work makes use of association rules to discard classes not playing a role according to some software metrics, so ML is not directly responsible for the detection~\citep{gueheneuc2004}. Similarly, decision trees and neural networks were trained with a set of manually defined features --named predictors-- to reduce the number of false positives detected after a structural analysis of the code~\citep{ferenc2005}. A collection of software metrics was used to feed a neural network, whose goal is the identification of classes potentially playing a role in the DP before addressing the detection~\citep{uchiyama2011}.

The rest of methods rely on ML techniques to make a decision about the presence of a DP in the code. \cite{alhusain2013} train multiple artificial neural networks, one for each role, with different subsets of metrics chosen via feature selection. They also proposed a second classifier to carry out the detection after filtering some candidates as in \citep{uchiyama2011}. In~\citep{chihada2015}, the classification model is created using support vector machines. This method uses as an input a labelled set of manually identified pattern implementations and a set of metrics associated with their roles. Then a subgraph isomorphic algorithm is executed for candidate design patterns to be extracted from the code. In MARPLE~\citep{zanoni2015}, software metrics are not taken as inputs but a set of structural and behavioural properties, such as object compositions or method delegations. Implemented as an Eclipse plugin, MARPLE makes use of several clustering and classification algorithms implemented by Weka.\footnote{\emph{Weka~3: Data Mining Software in Java}, available from \url{https://www.cs.waikato.ac.nz/ml/weka} (accessed June 11, 2020)} More particularly, MARPLE is able to use different highly interpretable classifiers like decision trees. However, it does not use DP instances directly as an input. Since it executes a clustering algorithm prior to detection, it makes the resulting models not as representative of the original samples as expected and, consequently, more difficult to understand.

More recently, DPD has been treated as a multi-classification problem, where the goal is to determine which specific pattern is implemented by the input code~\citep{Dwivedi2018}. Here, three different black-box classifiers --- neural networks, support-vector machines and random forests --- are trained using up to 67 software metrics. Convolutional neural networks and random forest have been also applied to learn from feature maps~\citep{thaller2019}. These elements are defined based on the occurrences of certain microstructures in the code.

Our proposal, \ourtechnique, belongs to the area of ML-based techniques for DPD. Compared to non-ML-based approaches, \ourtechnique does not need prior knowledge about the properties of the DP to be detected. Therefore, it can be adapted to organisational policies and different programming styles. As other ML-based methods, it automatically discovers which are the  properties that best define each DP by exploring previous implementations from code repositories. As a counterpart, ML-based approaches require a labelled set of implementations, which may not be available in all contexts. Focusing on other ML-based alternatives, \ourtechnique differs from the existing approaches in the combination of software metrics and properties as input for learning. As for the algorithm, it uses G3P to produce a rule-based classifier instead of black-box detectors, such as neural networks or support vector machines. Outcomes from these kinds of models are harder to understand, which also reduces their likelihood for adoption by practitioners~\citep{Rana14}.

\section{The \ourtechnique approach} \label{sec:approach}

\begin{figure}[!t]
\centerline{\includegraphics[scale=0.65]{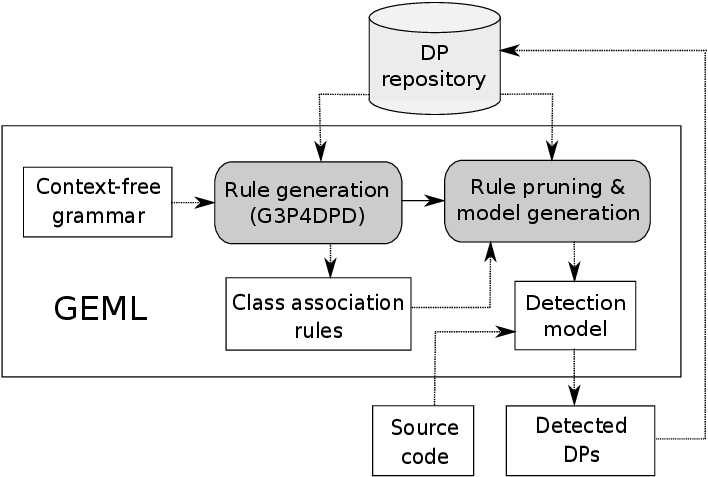}}
\caption{Two-phased model for design pattern detection}
\label{fig:model}
\end{figure}

Fig.~\ref{fig:model} depicts the overall structure of \ourtechnique. Taking the approach as a black-box, the inputs required consist in the organisational repository containing patterns predicted in the past (\ie samples), and the source code for which the detection process is carried out. After the execution of \ourtechnique, the detected DPs are returned. These patterns can be incorporated to the organisational repository for future detections, so that the detection model is progressively adapted to the specific development culture. Going into the detail of our approach, \ourtechnique has been proposed as a two-phased model. Firstly, the set of both structural, behavioural and metric-based software properties that best describe the design pattern under analysis are learned in form of rules. With this aim, a representative set of labelled DP implementations are scrutinised from the code repository, which contains both positive and negative samples. Keeping negative samples could provide relevant information about realistic scenarios, since similarities to a real DP could lead to misclassification otherwise. In fact, a high number of negative samples benefits the robustness of the detection process, specially when the amount of positive examples in public repositories is low~\citep{alhusain2013,thaller2019}. Every pattern instance within the repository is characterised by its constituent elements, its source code and the role mapping, \ie the correspondence between elements and roles.

The set of class association rules must be compliant with the CFG, formally defining the language syntax required to state the properties and constraints of the pattern. The G3P algorithm for DPD (G3P4DPD) has been developed to search for those rules that best characterise the implementations available in the repository. Since the resulting CARs have a descriptive nature, they could not be directly used for detection purposes. During the second step of the process, the most effective rules for the construction of the detection model will be chosen. With this aim, a pruning method is applied first to obtain the minimum set containing the most descriptive rules. Then, a strategy is followed to decide how to arrange the resulting rule set in order to build the detector, according to the precepts of different methods for associative classification~\citep{thabtah2007}. Both phases are explained in detail next.

\subsection{G3P-based Algorithm for Rule Generation} \label{subsec:g3p4dpd}

Algorithm~\ref{alg:g3p4dpdAlg} shows the general procedure of G3P4DPD, which receives five inputs: the number of generations (\textit{maxGen}), the population size (\textit{popSize}), the number of individuals or CARs to be returned  (\textit{extPopSize}), the grammar (\textit{cfg}) and the source code repository (\textit{repo}). G3P4DPD keeps an external archive (\textit{extPop}) composed of the most accurate individuals according to their evaluation. They are returned as output.

\begin{algorithm}[!t]
\SetKwInOut{Input}{Input}
    \SetKwInOut{Output}{Output}
    \Input{maxGen, popSize, extPopSize, cfg, repo}
    \Output{extPop}
    pop $\leftarrow$ generateRules(popSize, cfg) \\
    extPop $\leftarrow$ $\emptyset$ \\
    evaluate(pop, repo) \\
    \While{generation $<$ maxGen}{
        parents $\leftarrow$ select(pop $\cup$ extPop, popSize) \\
        offspring $\leftarrow$ crossover(parents) \\       
        \eIf{random() $<$ 0.5}{
   			offspring $\leftarrow$ diversityMutator(offspring)\;
   		}
        {
   			offspring $\leftarrow$ dpdMutator(offspring)\;
  		}
        evaluate(offspring, repo) \\
        pop $\leftarrow$ offspring \\
        extPop $\leftarrow$ update(pop $\cup$ extPop, extPopSize) \\
        generation++
     }
    \caption{G3P4DPD}
    \label{alg:g3p4dpdAlg}
\end{algorithm}

Regarding its operation, G3P4DPD firstly creates a random population (\textit{pop}) of \textit{popSize} individuals conformant with the syntax prescribed by the grammar, \textit{cfg}. The external archive, \textit{extPop}, is then initialised to the empty set. Initial individuals of \textit{pop} are evaluated by computing the fitness function, which obtains the support of the rule calculated from the repository \textit{repo}. At this point, the algorithm iterates until the \textit{maxGen}-th generation is reached. For each generation, the selection operator returns \textit{popSize} individuals (parents) from the union between population \textit{pop} and population \textit{extPop}. Next, the crossover operator is applied with a certain probability over pairs of parents. After combining their genotypes, the obtained \textit{offspring} are then mutated. The choice of the specific mutator is made randomly --- between two possible operators --- and seeks for a balance between search space exploration and the type and magnitude of the changes applied. On the one hand, \textit{diversityMutator} looks for the improvement of population diversity by including new properties in the comparison expressions of the rule. On the other hand, \textit{dpdMutator} is a DPD-specific mutator that aims to generate rules describing both positive and negative samples. Once these new offspring have been evaluated in terms of their support, they all replace the current population. Finally, the external archive is updated to keep those rules exceeding a certain support and confidence threshold, while ensuring that it does not exceed \textit{extPopSize} elements, \ie rules, and they are not repeated or redundant. Next sections explain in further detail how solutions are encoded and how the aforementioned genetic operators operate.

\subsubsection{Encoding}\label{subsec:g3p4dpd_encoding}

\begin{figure}[!t]
    \centering
    \begin{subfigure}[!t]{0.45\textwidth}
		\includegraphics[scale=0.6]{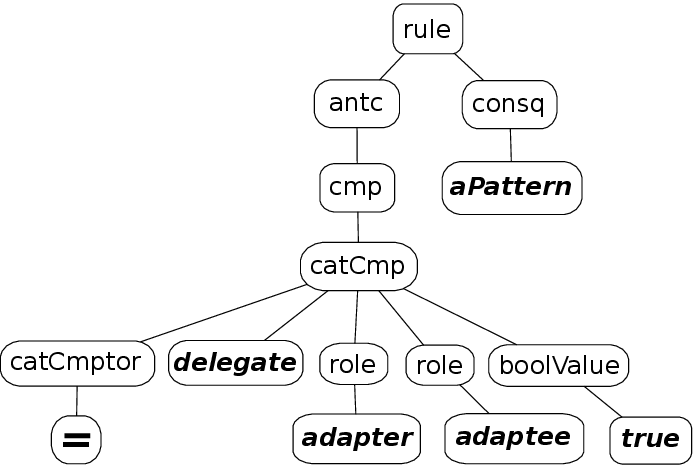}
        \caption{Genotype}
         \label{subfig:genotype}
    \end{subfigure}
    \par\bigskip
    \begin{subfigure}[!t]{0.45\textwidth}
        \includegraphics[scale=0.6]{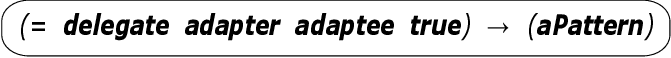}
        \caption{Phenotype}
        \label{subfig:phenotype}
    \end{subfigure}
    \caption{Example of correspondence between genotype and phenotype for an illustrative individual}
    \label{fig:individual}
\end{figure}

On the one hand, the genotype of an individual is represented by a valid derivation tree of production rules, as formally defined by the CFG. On the other hand, its phenotype denotes the corresponding class association rule. As an example, Fig.~\ref{fig:individual} shows a genotype/phenotype mapping, the described rule stating that, if the element playing the role \textit{adapter} delegates some functionality to the element \textit{adaptee}, then this would expected to be a valid Adapter pattern.

As explained in Section~\ref{sec:background_g3p}, any CFG requires the definition of sets of terminal and non-terminal symbols, and the production rules to derive valid expressions from a root, non-terminal symbol. In this context, terminal symbols represent the software metrics, as well as the structural and behavioural properties that allow identifying the presence of a pattern instance, whereas non-terminal symbols define the derivable elements, including all the components needed to build comparison expressions between the aforementioned properties. Production rules determine the derivation steps that lead to the generation of detection rules, which are ultimately written in terms of terminal symbols. Notice that the CFG could be certainly customised to a specific design pattern, thus allowing a more efficient description of the most relevant properties of that pattern.

\begin{figure}[!t]
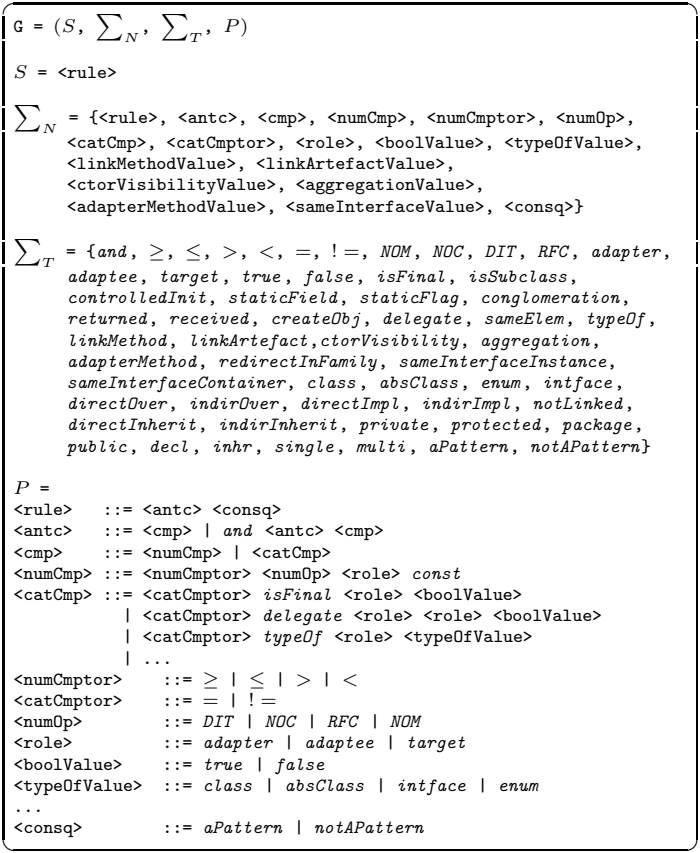

\begin{center}
\begin{lstlisting}[xleftmargin=0.0\textwidth, xrightmargin=0.0\textwidth, basicstyle=\scriptsize\ttfamily,]
G = (*@$(S$@*), (*@$\sum_N$@*), (*@$\sum_T$@*), (*@$P)$@*)

(*@$S$@*) = <rule>

(*@$\sum_N$@*) = {(*@<rule>,@*) (*@<antc>,@*) (*@<cmp>,@*) (*@<numCmp>,@*) (*@<numCmptor>,@*) (*@<numOp>,@*) (*@<catCmp>,@*) (*@<catCmptor>,@*) (*@<role>,@*) (*@<boolValue>,@*) (*@<typeOfValue>,@*) (*@<linkMethodValue>,@*) (*@<linkArtefactValue>,@*) (*@<ctorVisibilityValue>,@*) (*@<aggregationValue>,@*) (*@<adapterMethodValue>,@*) (*@<sameInterfaceValue>,@*) (*@<consq>@*)}

(*@$\sum_T$@*) = {(*@\textit{and}@*), (*@\textit{$\geq$}@*), (*@\textit{$\leq$}@*), (*@\textit{$>$}@*), (*@\textit{$<$}@*), (*@\textit{$=$}@*), (*@\textit{$!=$}@*), (*@\textit{NOM}@*), (*@\textit{NOC}@*), (*@\textit{DIT}@*), (*@\textit{RFC}@*), (*@\textit{adapter}@*), (*@\textit{adaptee}@*), (*@\textit{target}@*), (*@\textit{true}@*), (*@\textit{false}@*), (*@\textit{isFinal}@*), (*@\textit{isSubclass}@*), (*@\textit{controlledInit}@*), (*@\textit{staticField}@*), (*@\textit{staticFlag}@*), (*@\textit{conglomeration}@*), (*@\textit{returned}@*), (*@\textit{received}@*), (*@\textit{createObj}@*), (*@\textit{delegate}@*), (*@\textit{sameElem}@*), (*@\textit{typeOf}@*), (*@\textit{linkMethod}@*), (*@\textit{linkArtefact}@*),(*@\textit{ctorVisibility}@*), (*@\textit{aggregation}@*), (*@\textit{adapterMethod}@*), (*@\textit{redirectInFamily}@*), (*@\textit{sameInterfaceInstance}@*), (*@\textit{sameInterfaceContainer}@*), (*@\textit{class}@*), (*@\textit{absClass}@*), (*@\textit{enum}@*), (*@\textit{intface}@*), (*@\textit{directOver}@*), (*@\textit{indirOver}@*), (*@\textit{directImpl}@*), (*@\textit{indirImpl}@*), (*@\textit{notLinked}@*), (*@\textit{directInherit}@*), (*@\textit{indirInherit}@*), (*@\textit{private}@*), (*@\textit{protected}@*), (*@\textit{package}@*), (*@\textit{public}@*), (*@\textit{decl}@*), (*@\textit{inhr}@*), (*@\textit{single}@*), (*@\textit{multi}@*), (*@\textit{aPattern}@*), (*@\textit{notAPattern}@*)}

(*@$P$@*) =
<rule>   ::= <antc> <consq>
<antc>   ::= <cmp> | (*@\textit{and}@*) <antc> <cmp>
<cmp>    ::= <numCmp> | <catCmp>    
<numCmp> ::= <numCmptor> <numOp> <role> (*@\textit{const}@*)
<catCmp> ::= <catCmptor> (*@\textit{isFinal}@*) <role> <boolValue>
           | <catCmptor> (*@\textit{delegate}@*) <role> <role> <boolValue>
           | <catCmptor> (*@\textit{typeOf}@*) <role> <typeOfValue>
           | ...
<numCmptor>    ::= (*@$\geq$@*) | (*@$\leq$@*) | (*@$>$@*) | (*@$<$@*)
<catCmptor>    ::= (*@$=$@*) | (*@$!=$@*)
<numOp>        ::= (*@\textit{DIT}@*) | (*@\textit{NOC}@*) | (*@\textit{RFC}@*) | (*@\textit{NOM}@*)
<role>         ::= (*@\textit{adapter}@*) | (*@\textit{adaptee}@*) | (*@\textit{target}@*)
<boolValue>    ::= (*@\textit{true}@*) | (*@\textit{false}@*)
<typeOfValue>  ::= (*@\textit{class}@*) | (*@\textit{absClass}@*) | (*@\textit{intface}@*) | (*@\textit{enum}@*)
...
<consq>        ::= (*@\textit{aPattern}@*) | (*@\textit{notAPattern}@*)
\end{lstlisting}
\end{center}
\caption{Grammar used by the G3P4DPD algorithm}
\label{fig:grammar}
\end{figure}

Fig.~\ref{fig:grammar} shows the proposed grammar for the DPD process. Some production rules were omitted for readability. The first production rule listed in $P$ determines that the root symbol, \textit{$<$rule$>$}, can be derived into two other non-terminal symbols, \textit{$<$antc$>$} and \textit{$<$consq$>$}, representing the antecedent and the consequent, respectively. On the one hand, the antecedent can be defined as an inclusive sequence of numerical (\textit{$<$numCmp$>$}) and categorical (\textit{$<$catCmp$>$}) comparisons. On the other hand, \textit{$<$consq$>$} can be derived into two possible terminals, expressing whether it is a positive detection of a design pattern instance, \textit{aPattern}, or not, \textit{notAPattern}. In addition, it is worth noting that this grammar has been slightly customised for the Adapter pattern, the non-terminal \textit{$<$role$>$} being derived into \textit{adapter}, \textit{adaptee} and \textit{target}. Similarly, adaptations of the grammar to other patterns could be also made straightforward by just setting the operators and roles that are more relevant to each pattern from those available, or exceptionally by implementing new operators. A more general possibility is to make the complete list of operators available to the algorithm and let it ``find'' which are the most relevant for the search. In any case, G3P4DPD would not be affected, since the algorithm receives the grammar definition as a parameter and it will adapt the search in consequence.

As mentioned above, each property of the DP is declared by the antecedent of the rule as a comparison, written as an expression in prefix form containing a comparator, an operator, one or more arguments, and a value. Numerical comparisons, \textit{numCmp}, allow the representation of those properties based on software metrics. Each comparison comprises a numerical comparator ($<$, $>$, $\leq$ or $\geq$); a numerical operator that computes an object-oriented software metric from the CK (Chidamber--Kemerer) metric suite~\citep{chidamber1994}, such as NOM (number of methods), NOC (number of children), DIT (depth of inheritance tree) or RFC (response for a class); an argument referring to the measured element playing a role in the pattern; and a numerical constant (\textit{const}). As an example, the comparison ``\texttt{NOC(target)$<$1}'' indicates that the \textit{target} class has no subclasses. 
Likewise, other numerical software metrics could be implemented and added to the grammar without recompiling G3P4DPD. 

On the other hand, categorical comparisons involve structural and behavioural properties. Structural properties serve to express characteristics of structural elements, such as a class being abstract or final, as well as the relationships between these elements, such as an aggregation or generalisation. Behavioural properties refer to those interactions between classes or method invocations that can be analysed from static code. These comparisons are composed by a categorical comparator (\textit{=} or \textit{!=}); a categorical operator like \textit{linkArtefact} or \textit{delegate}; a number of arguments receiving the participating roles; and a value. As an example of a categorical comparison, ``\texttt{typeOf(target)=intface}'' describes that the \textit{target} role is implemented by an interface. Table~\ref{tab:operators} shows the full list of available operators, their signatures and description. It is worth noting that categorical operators are mainly based on elemental design patterns~\citep{smith2012}, micro patterns~\citep{gil2005} and design patterns clues~\citep{fontana2011}. They provide language elements that can be easily understood by the software engineer.

\begin{table*}[!t]
\caption{Grammar operators to describe design pattern implementations}
\centering
\small{
\begin{tabular}{| c | l | l |}
\hline

\multicolumn{3}{|c|}{\raggedright \textbf{Numerical operators}} \\ \hline
\textbf{Signature}			& \multicolumn{2}{c|}{\raggedright \textbf{Description}} \\ \hline

\textit{NOM($r_1$)}		& \multicolumn{2}{p{0.7\textwidth}|}{\raggedright Number of methods of $r_1$} \\ \hline

\textit{NOC($r_1$)}		& \multicolumn{2}{p{0.7\textwidth}|}{\raggedright Number of children directly inherited from $r_1$} \\ \hline

\textit{DIT($r_1$)}		& \multicolumn{2}{p{0.7\textwidth}|}{\raggedright Depth of the inheritance tree from $r_1$} \\ \hline

\textit{RFC($r_1$)}		& \multicolumn{2}{p{0.7\textwidth}|}{\raggedright Number of distinct methods and constructors potentially invoked when an object of $r_1$ receives a message} \\ \hline

\multicolumn{3}{|c|}{\raggedright \textbf{Categorical operators}} \\ \hline
\textbf{Signature}			& \multicolumn{2}{c|}{\raggedright \textbf{Description}} \\ \hline

\textit{isFinal($r_1$)}			& \multicolumn{2}{p{0.7\textwidth}|}{\raggedright \textit{true} if $r_1$ cannot be extended; \textit{false} otherwise} \\ \hline

\textit{isSubclass($r_1$)}		& \multicolumn{2}{p{0.7\textwidth}|}{\raggedright \textit{true} if $r_1$ is a subclass; \textit{false} otherwise} \\ \hline

\textit{controlledInit($r_1$)}	& \multicolumn{2}{p{0.7\textwidth}|}{\raggedright \textit{true} if $r_1$ instantiates itself within an \textit{if} or \textit{while} block; \textit{false} otherwise} \\ \hline

\textit{controlledExcept($r_1$)}	& \multicolumn{2}{p{0.7\textwidth}|}{\raggedright \textit{true} if $r_1$ uses exceptions and an static flag to control its instantiation; false otherwise} \\ \hline

\textit{conglomeration($r_1$)}		& \multicolumn{2}{p{0.7\textwidth}|}{\raggedright \textit{true} if 2 or more methods of $r_1$ are invoked from another method from $r_1$; \textit{false} otherwise} \\ \hline

\textit{returns($r_1$,$r_2$)}			& \multicolumn{2}{p{0.7\textwidth}|}{\raggedright \textit{true} if a value of type $r_2$ is returned from a method of $r_1$; \textit{false} otherwise} \\ \hline

\textit{receives($r_1$,$r_2$)}			& \multicolumn{2}{p{0.7\textwidth}|}{\raggedright \textit{true} if a method of $r_2$ receives a value of type $r_1$ as argument; \textit{false} otherwise} \\ \hline

\textit{createObj($r_1$,$r_2$)}		& \multicolumn{2}{p{0.7\textwidth}|}{\raggedright \textit{true} if $r_1$ instantiates $r_2$; \textit{false} otherwise} \\ \hline

\textit{delegates($r_1$,$r_2$)}		& \multicolumn{2}{p{0.7\textwidth}|}{\raggedright \textit{true} if a method of $r_1$ invokes a method of $r_2$; \textit{false} otherwise} \\ \hline

\textit{sameElem($r_1$,$r_2$)}			& \multicolumn{2}{p{0.7\textwidth}|}{\raggedright \textit{true} if $r_1$ and $r_2$ are coded by the same artefact; \textit{false} otherwise} \\ \hline

\textit{typeOf($r_1$)}						& \multicolumn{2}{p{0.7\textwidth}|}{\raggedright Returns the type of the artefact implementing $r_1$ (\textit{absClass}, \textit{intface}, \textit{enum}, \textit{class})} \\ \hline

\textit{linkMethod($r_1$,$r_2$)}			& \multicolumn{2}{p{0.7\textwidth}|}{\raggedright Indicates if a method of $r_1$ directly or indirectly overrides (\textit{directOver}, \textit{indirOver}), implements (\textit{directImpl}, \textit{indirImpl}) or is \textit{notLinked} to a method of $r_2$} \\ \hline

\textit{linkArtefact($r_1$,$r_2$)}				& \multicolumn{2}{p{0.7\textwidth}|}{\raggedright Returns the sort of link between the artefacts playing $r_1$ and $r_2$ (\textit{directInherit}, \textit{indirInherit}, \textit{directImpl}, \textit{indirImpl}, \textit{notLinked})} \\ \hline

\textit{ctorVisibility($r_1$)}					& \multicolumn{2}{p{0.7\textwidth}|}{\raggedright Returns the visibility of the less restrictive constructor of $r_1$ (\textit{private}, \textit{protected}, \textit{package}, \textit{public})} \\ \hline

\textit{aggregation($r_1$,$r_2$)}				& \multicolumn{2}{p{0.7\textwidth}|}{\raggedright Returns information about an attribute of $r_2$ declared in $r_1$ in terms of its visibility and instantiability
} \\ \hline

\textit{adapterMethod($r_1$,$r_2$,$r_3$)}	& \multicolumn{2}{p{0.7\textwidth}|}{\raggedright Returns if a declared (\textit{decl}) or inherited (\textit{inhr}) method of $r_1$, implemented from $r_3$, delegates in a method of $r_2$; \textit{notLinked} otherwise} \\ \hline

\textit{redirectInFamily($r_1$)}	& \multicolumn{2}{p{0.7\textwidth}|}{\raggedright Returns if a declared method of $r_1$ is delegated in a class or interface being extended or implemented by $r_1$, once (\textit{single}) or multiple times (\textit{multi}); \textit{notLinked} otherwise} \\ \hline

\textit{sameInterfaceInstance($r_1$,$r_2$)}	& \multicolumn{2}{p{0.7\textwidth}|}{\raggedright Returns if $r2$ is a class or interface being extended or implemented by $r_1$, which has one (\textit{single}) or multiple (\textit{multi}) fields of a class or interface being extended or implemented by $r_2$; \textit{notLinked} otherwise} \\ \hline

\textit{sameInterfaceContainer($r_1$,$r_2$)}	& \multicolumn{2}{p{0.7\textwidth}|}{\raggedright \textit{true} if $r2$ is a class or interface being extended or implemented by $r_1$, which defines a collection of a class or interface being extended or implemented by $r_2$; \textit{false} otherwise} \\ \hline

\end{tabular}
}
\label{tab:operators}
\end{table*}

These structures can be expressed as categorical operators checking whether a code artefact satisfies a property on a true/false basis. For instance, \textit{Abstract Interface} is an elemental design pattern checking whether a given code artefact is an interface. Usually, an artefact is a class or an interface. However, it largely depends on the specific programming language, and other artefacts could be considered like enumerations, or subtypes of artefacts, such as abstract and concrete classes. It implies that a large number of operators like \textit{isConcrete}, \textit{isAbstract}, \textit{isInterface} and \textit{isEnumeration} would be necessary to cover all types of artefacts. Besides, these operators would be strongly correlated, \eg if \textit{isConcrete} is \textit{true}, then the others should be \textit{false}, what could make rules redundant and add noise to the learning process. Therefore, the CFG declares multi-valued categorical operators, which are able to return different alternative terms instead. This is the case of \textit{typeOf}, which allows analysing a source code artefact and returns a single value for all these possibilities. In this way, this operator replaces a number of redundant and unnecessarily inefficient boolean operators, while boosts the generation of more compact and readable rules.

\subsubsection{Genetic Operators}\label{subsubsec:g3p4dpd_operators} 

There are three genetic operators to be considered in this evolutionary schema: selection, crossover and mutation. The selection operator is applied at the beginning of each generation in order to choose a set of parents for breeding. With this aim, a binary tournament selects two individuals from the union of \textit{pop} and \textit{extPop}, taking the best according to their fitness. This process is repeated until \textit{popSize} parents are selected.

\begin{figure*}[!t]
\centerline{\includegraphics[scale=0.905]{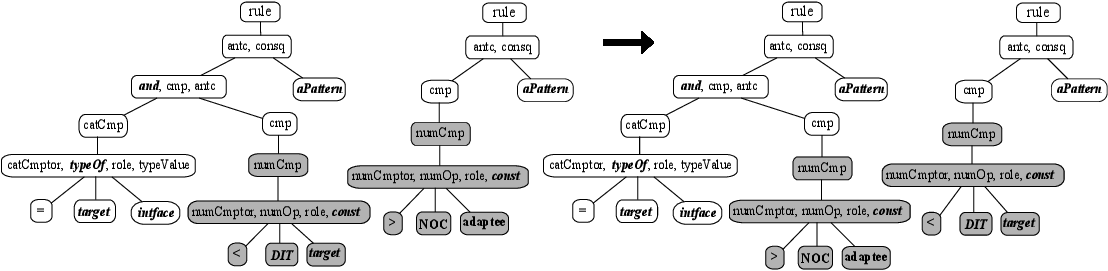}}
\caption{Example of the crossover operator}
\label{fig:crossover}
\end{figure*}

Regarding the crossover, this operator is applied to each pair of parents according to a given probability. The operator works by randomly selecting and swapping a single comparison from their respective genotype. Finding such a comparison requires traversing the derivation tree in pre-order until a \textit{$<$cmp$>$} symbol is reached, delimited by \textit{and}, \textit{$<$cmp$>$} or \textit{$<$consq$>$}, as illustrated in Fig.~\ref{fig:crossover}. Then, each offspring is mutated.

\begin{figure}[!t]
    \centering
    \begin{subfigure}[!t]{0.5\textwidth}
		\includegraphics[scale=1.25]{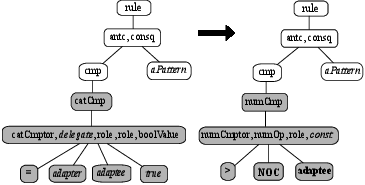}
        \caption{Operator \textit{diversityMutator}}
         \label{subfig:diversityMutator}
    \end{subfigure}
    \par\bigskip
    \begin{subfigure}[!t]{0.5\textwidth}
        \includegraphics[scale=1.25]{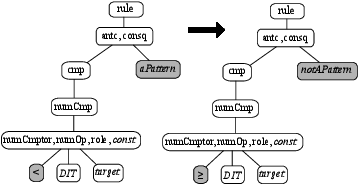}
        \caption{Operator \textit{dpdMutator}}
        \label{subfig:dpdMutator}
    \end{subfigure}
    \caption{Examples of the mutation operators}
    \label{fig:mutation}
\end{figure}

Two possible variants of mutator have been considered for this problem. Firstly, the so-called \textit{diversityMutator} aims at generating rules with novel comparisons. To this end, it selects a number of comparisons and rebuilds the individual by randomly deriving these \textit{$<$cmp$>$} until new terminal symbols are obtained. The number of comparisons is chosen with a random roulette wheel that promotes small changes against those significantly altering the individual. As an example, Fig.~\ref{subfig:diversityMutator} shows how a categorical comparison is replaced by a numerical comparison. Secondly, \textit{dpdMutator} aims at obtaining rules describing both positive and negative samples. The rationale behind this operator is that if a comparison describes the correct implementation of a pattern, its negation should imply an incorrect result. This operator traverses the derivation tree and rewrites the logical meaning of each comparison with an inverse probability to the number of comparisons in the antecedent. With this aim, it simply switches the comparison symbol, \eg $<$ is replaced by $\geq$, as illustrated in Fig.~\ref{subfig:dpdMutator}. In addition, the terminal symbol of the consequent, \ie the class label, will be also inverted with a probability of 0.5. For instance, if the comparison ``\texttt{typeOf(target)=intface}'' describes a recurrent property of positive samples for the Adapter pattern, ``\texttt{typeOf(target)!=intface}'' is likely to represent negative samples. At the end of the mutation operation, regardless of the specific variant being used, the mutated offspring is returned.

\subsection{Construction of the Detection Model}\label{subsec:classification}

After completing the first step explained above, the G3PDPD algorithm returns its external archive composed of CARs. However, they still require to be scrutinised to select strictly those that will be part of the detection model. With this aim, the database coverage method is computed on the complete set of rules. This pruning method was originally defined for CBA~\citep{liu1998}, a general approach for associative classification, and later applied by other well-known algorithms like CMAR~\citep{li2001} or CPAR~\citep{yin2003}.

Algorithm~\ref{alg:dCoverage} shows the pseudocode of the pruning method, conveniently adapted to this problem. The procedure takes the archive returned by G3P4DPD, \textit{extPop}, the repository, \textit{repo}, and a coverage \textit{threshold}. As an output, it returns the pruned set of rules, \textit{ruleSet}. To do this, rules are sorted according to their confidence, support and size, \ie number of comparisons, respectively. More precisely, given two rules \textit{$R_1$} and \textit{$R_2$}, it is said that \textit{$R_1$} \textit{precedes} \textit{$R_2$} iff any of these conditions is satisfied: ($a$) conf(\textit{$R_1$})~$>$~conf(\textit{$R_2$}); ($b$) conf(\textit{$R_1$})~$=$~conf(\textit{$R_2$}) and supp(\textit{$R_1$})~$>$~supp(\textit{$R_2$}); or ($c$) conf(\textit{$R_1$})~$=$~conf(\textit{$R_2$}), supp(\textit{$R_1$})~$=$~supp(\textit{$R_2$}) and \textit{\#comparison}($R_1$)~$<$~\textit{\#comparison}($R_2$). Then, all repository samples are scanned for each rule searching for those matching the antecedent. When positive, they are conveniently added to the list \textit{covSamples}. In addition, if the rule correctly classifies at least one of these samples, \ie the flag \textit{marked} is true, then it will be added to \textit{ruleSet} and the counter of covered samples in \textit{repo} will be increased. In case a sample is covered by a number of rules greater than or equal to the \textit{threshold}, it will be removed from \textit{repo} so that this sample does not need to be checked again. Finally, if the rule does not contribute to increase the current detection capability, it will be discarded. This happens when the rule detects the same samples as other rules with higher confidence and support. This process is repeated until all samples in \textit{repo} have been covered or until there are no more rules left.

\begin{algorithm}[!t]
\SetKwInOut{Input}{Input}
    \SetKwInOut{Output}{Output}
    \Input{extPop, repo, threshold}
    \Output{ruleSet}
    ruleSet $\leftarrow$ $\emptyset$ \\
    sort(extPop) \\
    \ForEach{rule \textbf{in} extPop}{
        marked $\leftarrow$ false \\
        covSamples $\leftarrow$ $\emptyset$ \\
        \ForEach{sample \textbf{in} repo}{
           \If{match(rule, sample)}{
                covSamples $\leftarrow$ covSamples $\cup$ sample \\
       			\If{rule \textit{correctly classifies} sample} {
       			    marked $\leftarrow$ true \\
       			}
       		}
  		}
  		\If{marked}{
  		    ruleSet $\leftarrow$ ruleSet $\cup$ rule \\
            increaseCoverage(repo, covSamples) \\
  		    removeCoveredSamples(repo, threshold) \\
  		}
        \If{empty(repo)}{
  		    \textbf{break} \\
  		}
     }
    \caption{Database coverage}
    \label{alg:dCoverage}
\end{algorithm}

After the pruning procedure, the selected rules need to be arranged in order to constitute the classifier. Here, the four most commonly referenced strategies from the field of associative classification have been considered for comparison~\citep{thabtah2007}: maximum likelihood (MAXL); dominant factor multiple label (DFML); DFML, as defined by CMAR (DFML$_{\chi^2}$); and DFML, as determined by CPAR (DFML$_{Lap}$). They all operate differently when a new, unknown sample is received. For instance, the MAXL method selects the highest ranked rule covering the new sample, its consequent implying the prediction. This approach has been criticised by some authors since a single rule becomes responsible for the decision of the classifier~\citep{li2001}. In contrast, DFML selects all rules covering the incoming sample and distributes them according to their consequent~\citep{hadi2016}. The partition containing more rules determines the class. This method has been adapted in different proposals by changing how the most representative partition is obtained. A well-known variant is DFML$_{\chi^2}$, where the weighted $\chi^2$ is computed for each partition, the set of rules with the highest value being returned. This measure analyses the strength of the rules based on their support and class frequency. Similarly, DFML$_{Lap}$ calculates the \textit{Laplace} accuracy to estimate the expected accuracy for each rule and selects the partition with its highest average. It should be noted that DFML$_{Lap}$ considers only the best $k$ rules within each partition, $k$ being a prefixed parameter. Then, after sorting the rules, the classifier is ready to receive new code, apply the rules and predict the presence of a DP implementation.


\section{Experimental Settings}\label{sec:settings}

We first pose the research questions that set out the objectives to be validated in this section. Then \new{we present the data repositories and} explain the methodology followed in the experiments. \new{The last section details} the experimental framework.

\subsection{Research Questions}\label{subsec:rquestions}

The conducted experiments aim at responding the following research questions (RQs):

\begin{itemize}
    \item \textit{RQ1. What configuration best informs our technique for each DP?} The combination of parameters for the G3P-based machine learning algorithm and the pruning method might lead to vast number of results. It would be convenient to objectively provide the best parameter setup for each DP.
    
    \item \textit{RQ2. For the software engineer, which are the most significant design elements and metrics to be considered for the detection of a given pattern?} The combination of numerical (metrics) and categorical (behavioural and structural properties) attributes offers the engineer the possibility of using a wide range of microstructures and operators at his/her convenience. In this way, after analysing which operators are most frequently used to describe each pattern, the experimentation can be replicated by adjusting them in order to find out whether the results maintain the detection performance.
    
    \item \textit{RQ3. Does \ourtechnique provide as much effectiveness as other ML-based DPD techniques?} It would be convenient to study if, taking a single configuration for all DPs, our model is still competitive against a similar ML-based approach. Reaching a general-purpose configuration would relieve the software engineer of the need to set up the model.
    
    \item \textit{RQ4. How does training conditions influence \ourtechnique?} As any other ML-based approach, \ourtechnique requires the availability of DP examples to learn from. Therefore, it is important to study how \ourtechnique behaves when few examples are provided. This would allow the software engineer to confirm the potential of the technique under different execution scenarios.
    
    \item \textit{RQ5. Does \ourtechnique provide as much effectiveness as other non-ML-based DPD techniques?} Similarly to RQ3, it would be interesting to analyse the benefits that \ourtechnique brings compared to other non-ML-based DPD tools currently available to the software engineer, thus ensuring that it also contributes to advance in the state of the practice.
    
\end{itemize}

\subsection{\new{Experimental Data Sources}}
\label{subsec:datasets}

The experimentation considers a total of 15 design patterns, covering the three different categories defined by~\citep{gamma1995}: creational (Abstract factory, Factory method and Singleton), behavioural (Command, Iterator, Observer, State, Strategy, Template Method and Visitor) and structural (Adapter, Bridge, Composite, Decorator and Proxy). These patterns provide different levels of complexity, as they differ in the number of roles and may present elements playing multiple roles. 

Samples of these design patterns have been extracted from public repositories, but other institutional data sources could be valid too. \new{More specifically, we conduct our experiments with implementations from two repositories, namely DPB and P-Mart, each with a different purpose. Firstly}, we consider DPB~\citep{dpb2012} \new{because} it provides the highest number of samples from publicly available repositories. Moreover, DPB was created by the authors of MARPLE to validate their approach, what ensures a fair comparison against this particular proposal, and contains both positive and negative samples, a requirement when experimenting with ML-based methods. DPB collects samples from nine industrial Java projects, plus one more project adopted from the literature. Table~\ref{tab:projects} lists the number of artefacts, methods, attributes and lines of code (LOC) of each software project contained by DPB. A limitation of this repository is that it only supports five design patterns (see Table~\ref{tab:candidates}, which includes the total number of samples [S], divided into positive [+] and negative [-]).

\new{In order to compare GEML against other non-ML-based proposals, we rely on P-Mart~\citep{gueheneuc2007p}, a peer-validated repository frequently used in other DPD studies. P-Mart includes samples of 20 design patterns distributed among all projects listed in Table~\ref{tab:projects}, with the exception of the DPExample project. Table~\ref{tab:samplesProjects} summarises the samples per project identified within P-Mart, as well as the number of existing DPs from those defined by Gamma. As can be observed, JHotDraw is the project containing the greatest variety of DPs (11), making it appropriate for its use as a test project when comparing GEML with other DPD studies, as will be explained later in Section~\ref{subsec:methodology}.}

\new{The use of P-Mart allows GEML to be set against other DPD methods reporting results for JHotDraw, but this project limits the number of DPs available for comparison. Notice that P-Mart was not originally designed to specifically experiment with ML-based approaches due to the reduced number of positive samples, so it has to be enlarged to be suitable for training~\citep{thaller2019}. To carry out a more complete, accurate and practicable analysis, it is important to provide GEML with a more extensive collection of DPs for training than those available in P-Mart. Therefore,} we have built a training repository in order to validate the applicability of \ourtechnique on \new{an even greater} number of design patterns. This is a common practice in DPD studies, especially in those presenting ML-based proposals~\citep{ferenc2005,alhusain2013,thaller2019}. Therefore, we have complemented the positive samples available in P-Mart with other instances manually validated from those recovered by DPD tools~\citep{ferenc2005,DeLucia2018}. In particular, we include those pieces of code for which both SSA and Ptidej reported the presence of a DP implementation with all its roles. In the case of Ptidej, this information is complete. In contrast, some samples of P-Mart are not fully labelled, which has caused us to double-check the repository to maintain only those instances that describe all their roles. As for the negative samples, a common approach is to generate random candidates from the code and select those that are more similar to positive samples according to some heuristics~\citep{thaller2019}. We follow a similar procedure, so that a maximum of three negative samples per positive sample have been generated.\footnote{Details of the procedure and the resulting training repository are available from the additional material (see page~\pageref{sec:additional-material}).} The final list of design patterns \new{available in the repository} is shown in Table~\ref{tab:candidates}, together with the number of training samples. After preliminary experiments, the number of positive samples for the remaining seven DPs available in P-Mart has proven to be insufficient for training.

\begin{table}[!t]
\caption{Properties of the Java projects used for experimentation}
\centering
\resizebox{0.475\textwidth}{!}{
\begin{tabular}{lrrrrrr}
\hline
Project             & Types & Methods & Attributes & LOC \\
\hline
DPExample           & 1749 & 4710 & 1786 & 32313 \\
QuickUML 2001       & 230 & 1082 & 421 & 9233 \\
Lexi v0.1.1 alpha   & 100 & 677 & 229 & 7101 \\
JRefactory v2.6.24  & 578 & 4883 & 902 & 79732 \\
Netbeans v1.0.x     & 6278 & 28568 & 7611 & 317542 \\
JUnit v3.7          & 104 & 648 & 138 & 4956 \\
JHotDraw v5.1       & 174 & 1316 & 331 & 8876 \\
MapperXML v1.9.7    & 257 & 2120 & 691 & 14928 \\
Nutch v0.4          & 335 & 1854 & 1309 & 23579 \\
PMD v1.8            & 519 & 3665 & 1463 & 41554 \\
\hline
\end{tabular}}
\label{tab:projects}
\end{table}

\begin{table}[!t]
\caption{Samples per design pattern in each experiment}
\centering
\resizebox{0.475\textwidth}{!}{
\begin{tabular}{|l|rrr|rrr|}
\hline
Design  & \multicolumn{3}{c|}{Experiment \#1} & \multicolumn{3}{c|}{Experiment \#2}\\\cline{2-7}
pattern  & + & - & S    & + & - & S\\
\hline
Adapter         & 618 & 603 & 1221      & 65 & 180 & 245\\
Fact. Method    & 562 & 482 & 1044      & 3 & 9 & 12\\
Decorator       & 93 & 154 & 247        & 2 & 6 & 8\\
Singleton       & 58 & 96 & 154         & 55 & 165 & 220\\
Composite       & 30 & 98 & 128         & 6 & 18 & 24\\
State           & - & - & -             & 147 & 412 & 559\\
Templ. Method   & - & - & -             & 50 & 147 & 197\\
Proxy           & - & - & -             & 19 & 57 & 76\\
Observer        & - & - & -             & 8 & 14 & 22\\
Strategy        & - & - & -             & 7 & 21 & 28\\
Abs. Factory    & - & - & -             & 6 & 18 & 24\\
Command         & - & - & -             & 5 & 15 & 20\\
Visitor         & - & - & -             & 3 & 9 & 12\\
Iterator        & - & - & -             & 3 & 9 & 12\\
Bridge          & - & - & -             & 2 & 6 & 8\\
\hline
\end{tabular}}
\label{tab:candidates}
\end{table}

\begin{table}[!t]
\caption{\new{Characteristics of the projects available in P-Mart and DPExample}}
\centering
\begin{tabular}{lrr}
\hline
Project & No. DPs & No. positive samples \\
\hline
DPExample           & 19 & 174\\
QuickUML 2001       & 6 & 7\\
Lexi v0.1.1 alpha   & 3 & 5\\
JRefactory v2.6.24  & 6 & 26\\
Netbeans v1.0.x     & 4 & 26\\
JUnit v3.7          & 5 & 8\\
JHotDraw v5.1       & 11 & 21\\
MapperXML v1.9.7    & 9 & 15\\
Nutch v0.4          & 8 & 15\\
PMD v1.8            & 9 & 14\\
\hline
\end{tabular}
\label{tab:samplesProjects}
\end{table}

\subsection{\new{Description of Experiments}}
\label{subsec:methodology}

\new{Three} experiments have been planned in order to find an answer to these RQs:

\begin{itemize}
    \item \textit{Experiment \#1}. In this experiment, we validate the detection capability of \ourtechnique under laboratory settings. More specifically, we study how the parameter tuning influences our proposal and analyse different configurations with respect to MARPLE~\citep{zanoni2015}, another method based on machine learning.
    
    \item \new{Experiment \#2. In this experiment we compare \ourtechnique against other recent non-ML-based DPD proposals: DePATOS~\citep{yu2018}, MLDA~\citep{Al-Obeidallah2019} and SparT~\citep{Xiong2019}. With this aim, for comparison purposes, we provide results on a P-Mart project, JHotDraw.}
    
    \item \textit{Experiment \#\new{3}}. This experiment focuses on more qualitative aspects, evaluating \ourtechnique in a more practical setting. \new{Here, GEML is tested on a project, DPExample, taken from a different repository, DPB, than the one used for training.} In this case, we compare our proposal with two non-ML-based DPD tools, SSA and Ptidej, which were run using the testing project as input. We choose these tools because they are the most frequently used for comparative purposes in DPD studies (selected in 88\% and 35\% of the works, respectively). In our preliminary analysis, we found other eight tools, but they were unavailable for download or failed in their installation or execution.
\end{itemize}

To cope with the intrinsic randomness of evolutionary algorithms, 30 independent runs with different random seeds are executed. For Experiment \#1, the possible bias due to the training partitions of the input repository \new{(DPB)} is solved by carrying out a stratified 10-fold cross validation in every execution. \new{In Experiment \#2, we set the conditions that allow the comparison with other DPD approaches on a widely studied project, JHotDraw. All the DPD under analysis provide access to the recovered instances in this project, allowing a fair comparison by inspecting their level of agreement with respect to P-Mart. Also, JHotDraw is the P-Mart project with a greater variety of DP implementations. The rest of projects within P-Mart are used as the training set. Here we only consider as truly DP samples those labelled as positive on P-Mart.} In Experiment \#\new{3} we want to simulate the process in which an engineer uses past projects to train the DPD model \new{over all the possible DPs}, and then apply the detection model on a new project. \new{Since none of the nine P-Mart projects contains samples of all the DPs, we used them together for training,} and the detection model is tested on DPExample. Since SSA and Ptidej do not require any training phase, they are directly executed using DPExample as input. True positive samples of DPExample have been manually revised.

Results are then reported in terms of the following commonly used classification measures: \textit{accuracy}, which indicates the percentage of instances correctly classified; \textit{recall}, which corresponds to the amount of DPs retrieved over the total number of samples within the repository; \textit{precision}, which measures how many DPs are positively detected as positive; \textit{specificity}, which indicates the proportion of negative samples correctly identified as negative; and \textit{$F_1$ score}, which calculates the harmonic mean between precision and recall. Once these performance metrics have been computed, the relevance of the results~\citep{Arcuri2014} needs to be statistically validated. Firstly, the Wilcoxon test allows performing pairwise comparisons, where the null hypothesis, $H_0$, hypothesises that both algorithms -- or configurations -- perform equally well. The distributions to be compared represent the results of 30 executions of the evolutionary algorithm for a given DP and performance measure. As multiple configurations are tested, p-values are conveniently adjusted by using the Holm's method. For those pairwise comparisons reporting significant differences, the Cliff's delta test is carried out to measure the effect size. A 95\% confidence level is considered for both tests.

\subsection{Experimental Framework}\label{subsec:framework}

An implementation of \ourtechnique is provided in Java using JCLEC~\citep{ventura2008jclec}, which includes functionalities and data structures to implement evolutionary algorithms. The VF2 algorithm~\citep{cordella2004}, available in the VFLib graph matching library,\footnote{\emph{VFLib}, available from \url{https://mivia.unisa.it/vflib/} (accessed June 20, 2020)} has been used for implementing the generation of candidates in Experiment \#2. Three additional libraries have been used: ckjm,\footnote{\textit{Chidamber and Kemerer Java Metrics (ckjm)}, available from \url{https://www.spinellis.gr/sw/ckjm} (accessed June 20, 2020)} an implementation of the CK metric suite; Java Parser,\footnote{\textit{JavaParser for processing Java code} available from \url{https://javaparser.org/} (accessed June 20, 2020)} a source code parser used to extract information and properties from code; and Javassist,\footnote{\textit{Javassist: Java bytecode engineering toolkit} available from \url{http://www.javassist.org/} (accessed June 20, 2020)} a similar library that takes bytecode as input. They both are used to implement the grammar operators during rule evaluation, although bytecode-based operators are preferred because they allow a faster computation of the properties.

Table~\ref{tab:exeCfg} lists the parameter setup and its variations for the detection model, including the G3P4DPD algorithm and the pruning procedure. The population size, the number of generations and the crossover probability have been set after preliminary experiments. The maximum number of derivations determines the limit of the genotype size, \ie how long a rule can be. This value has been set to 25, as larger values would hardly generate individuals with admissible support. In addition, the size of the external archive has not been restricted to avoid discarding interesting rules and to test the effectiveness of the pruning method. The rest of parameters will be set according to the conclusions extracted from the parameter study in Section~\ref{subsec:exp1_param}, which serves to determine their influence on the detection process. As for this analysis, support, confidence and coverage are configured according to the most frequently applied values within the AC field~\citep{liu1998,li2001}. 

\begin{table}[!t]
	\centering
    \caption{Parameter setup of the DPD model}
    \small{
	\begin{tabular}{c  c} 
		\hline
		Parameter & Value \\
		\hline
    Population size & 100 \\
		Number of generations & 150 \\
    Crossover probability & 0.8 \\
    Maximum number of derivations & 25 \\
	Coverage threshold & 1, 2, 3, 4 \\	
    Support threshold & 0.01, 0.05, 0.1 \\
    Confidence threshold & 0.5, 0.6, 0.7 \\
    \hline
	\end{tabular}
	}
	\label{tab:exeCfg}
\end{table}

\section{Experiment 1: Validation of the Detection Model}\label{sec:experiment1}

Experiment \#1 is explained in this section. The internal elements of the detection model are analysed by means of an extensive parameter study (RQ1). Then, we also focus on the selection of the grammar operators, revealing the most recurrent microstructures and design elements to describe each DP (RQ2). Finally, the results are discussed with special emphasis on the effectiveness of the proposal (RQ3).

\subsection{Parameter Study}\label{subsec:exp1_param}

In response to RQ1, we carry out a parameter study focused on the three influential aspects required for the most fitting parametrisation of \ourtechnique: the coverage threshold, which is a key parameter for rule pruning; support and confidence thresholds, which effect the production of high-quality rules during the evolutionary search; and the strategy selection, which determines how the pruned rules are arranged to form the DPD model (see the Additional Material and Appendix for complete results).

\subsubsection{Coverage Threshold}\label{subsubsec:exp1_param_full-analysis}

The coverage threshold determines how many rules a given sample should satisfy to be considered as covered by the detection model. The lower the threshold, the lower the number of resulting rules. Following the recommendations of the AC literature~\citep{liu1998,li2001}, we have selected four possible values, from 1 to 4. Default values for support and confidence have been set to 0.01 and 0.5. In addition, given that a classification strategy is required, MAXL has been chosen as the baseline procedure, since it is the simplest, only considering one rule for the classification. Again, the five classification measures are computed for each configuration, with special emphasis in $F_1$, which reflects the effect of both precision and recall. After 30 executions, the best average values for $F_1$ were obtained when the coverage threshold is equal to 1. As for the statistical analysis, the Wilcoxon test does not report significant differences in the case of the Adapter. In contrast, differences were found for the rest of patterns, especially when compared with largest values of the threshold. Considering these differences and the fact that lower coverage thresholds help reducing the size of the rule set, the threshold is set to 1 for all the patterns.

\subsubsection{Support, Confidence and Classification Strategy}\label{subsubsec:exp1_param_g3p}

The update mechanism of the archive is controlled by the support and confidence thresholds, looking for only preserving high-quality rules. In response to RQ1, we need to determine the parameters that return rules with the highest quality for each DP. With this aim, all the combinations of support and confidence thresholds (see Table~\ref{tab:exeCfg}), jointly with different classification strategies, have been analysed. The outcomes from these combinations can be found in the Appendix for each design pattern separately. Table~\ref{tab:resultsAllOperators} summarises the findings, showing the best combination of parameters with respect to $F_1$.

\begin{table*}[!t]
\caption{Best classification performance in terms of $F_1$ for each design pattern}
\label{tab:resultsAllOperators}
\centering
\scriptsize{
\begin{tabular}{c c c c c c c c }
\hline
 & Strategy & Supp-Conf & Accuracy & Precision & Recall & Specificity & $F_1$\\ 
 \hline
 Singleton            & DFML$_{\chi^2}$ & (0.01) - (0.7) & $ 0.9561 \pm 0.0136 $ & $ 0.9460 \pm 0.0147 $ & $ 0.9461 \pm 0.0298 $ & $ 0.9621 \pm 0.0118 $ & $ 0.9411 \pm 0.0202 $ \\
 Adapter              & DFML$_{\chi^2}$ & (0.01) - (0.7) & $ 0.8688 \pm 0.0022 $ & $ 0.8430 \pm 0.0028 $ & $ 0.9121 \pm 0.0032 $ & $ 0.8244 \pm 0.0038 $ & $ 0.8757 \pm 0.0020 $ \\
 Factory Method       & DFML$_{\chi^2}$ & (0.05) - (0.7) & $ 0.8304 \pm 0.0082 $ & $ 0.8113 \pm 0.0081 $ & $ 0.8965 \pm 0.0158 $ & $ 0.7533 \pm 0.0145 $ & $ 0.8503 \pm 0.0081 $ \\
 Decorator            & DFML$_{\chi^2}$ & (0.05) - (0.7) & $ 0.8229 \pm 0.0179 $ & $ 0.8043 \pm 0.0269 $ & $ 0.7251 \pm 0.0393 $ & $ 0.8824 \pm 0.0171 $ & $ 0.7501 \pm 0.0302 $ \\
 Composite            & DFML$_{\chi^2}$ & (0.05) - (0.6) & $ 0.8859 \pm 0.0181 $ & $ 0.7567 \pm 0.0408 $ & $ 0.8900 \pm 0.0700 $ & $ 0.8845 \pm 0.0243 $ & $ 0.7885 \pm 0.0425 $ \\ \hline
\end{tabular}
}
\end{table*} 

For the sake of clarity, results are mainly analysed based on the values obtained for $F_1$. In addition, notice that accuracy is not a fully reliable measure here, as long as the DPB repository is highly imbalanced. With respect to the classification performance of the different configurations, there is not a big difference for the Singleton, Adapter and Factory Method patterns. The same does not apply for the Decorator and Composite, however, for which differences of more than 20\% are reported. These differences mainly occur between those configurations using different classification strategies, the support and confidence thresholds having less effect.

Regarding the classification strategy, it is worth noting that the same set of CARs has been used as input for all the strategies. In the case of DFML$_{Lap}$, $k$ is set to 5, as suggested by its authors. Firstly, the results reveal that using only one rule for detection is not the best alternative, since MAXL is never able to return the best value for any performance measure. In fact, this strategy reaches significantly worse results for the Decorator and Composite patterns. Similarly, DFML$_{Lap}$ obtains low detection performance for these patterns, even worse than MAXL. Note that this strategy only takes into account a reduced number of rules for detection, which is given by the value of $k$. Thus, it could be argued that using a limited number of rules is not the best alternative for the DPD problem. Secondly, considering the strategies DFML and DFML$_{\chi^2}$, the latter stands out as the procedure that best exploits the detection capability of \ourtechnique. More specifically, the best values of $F_1$ are reached when using this strategy. Focusing on the negative samples, other strategies could obtain better specificity results but at the expense of lower recall values, meaning that less DP instances would be recovered. Therefore, only configurations using DFML$_{\chi^2}$ are considered when analysing how the support and confidence thresholds influence the detection model.

As for the support and confidence thresholds, the best performance is mostly reached for those configurations with a higher confidence threshold (0.7), the Composite being the only exception. We speculate that these rules are expected to provide more certainty in the detection. Nevertheless, notice that lower values are commonly used in the AC literature. Regarding the support, the Singleton and Adapter obtain the best results with the lowest support value (0.01), whilst 0.05 is the best value for the Factory Method, Decorator and Composite patterns. However, there is no best configuration with a support value of 0.1. Since this measure is related to the proportion of samples satisfying a rule, low values are needed to include those rules that are able to describe less common pattern implementations within the repository. Therefore, lower support values stand out as the best alternative, as suggested by the AC literature.

Finally, we summarise the main findings that give answer to RQ1:

\begin{itemize}
    \item Low support values are preferred for all design patterns, 0.01 showing better results for Singleton and Adapter. For Factory Method, Decorator and Composite, a support equal to 0.05 is recommended.
    \item A confidence threshold equal to 0.7 is a good general choice, as only Composite obtains better results when setting another value (0.6).
    \item For all design patterns, DFML$_{\chi^2}$ is the pruning strategy providing best performance. This strategy is especially beneficial for the Decorator and Composite patterns.
\end{itemize}

%
\subsection{Selection of Grammar Operators}\label{subsec:exp1_operators}

\begin{figure*}[!t]
    \centering
     \begin{subfigure}[!t]{0.49\textwidth}
		 \includegraphics[scale=0.49]{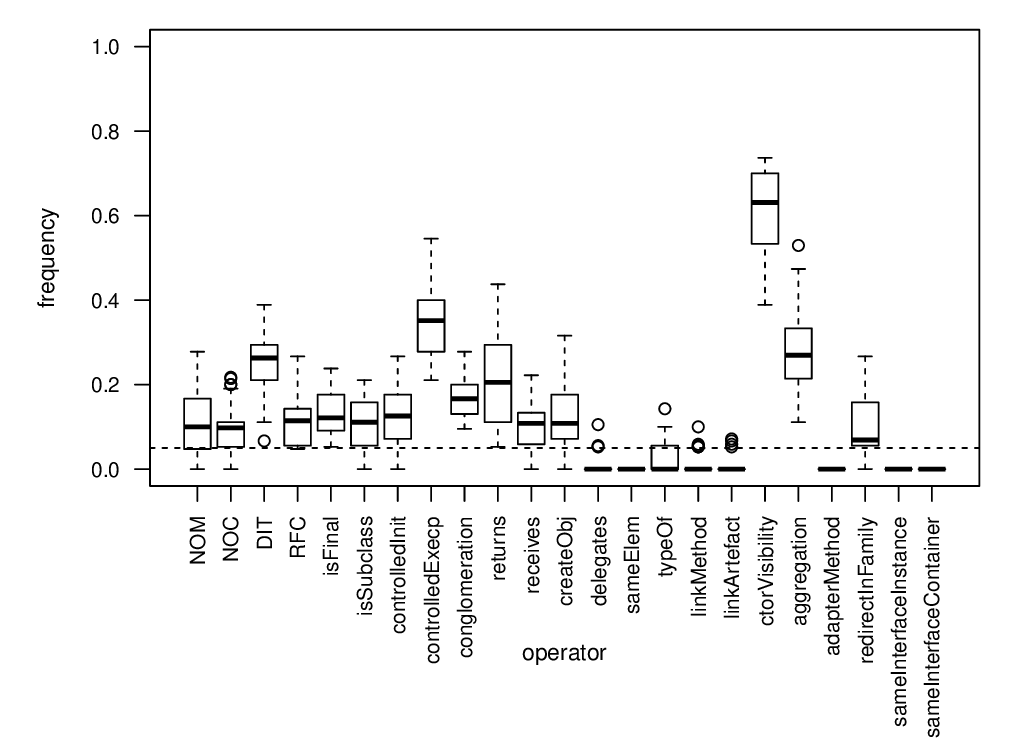}
        \caption{Singleton}
         \label{subfig:norm_singleton-op}
    \end{subfigure}
     \hfill
    \begin{subfigure}[!t]{0.49\textwidth}
        \includegraphics[scale=0.49]{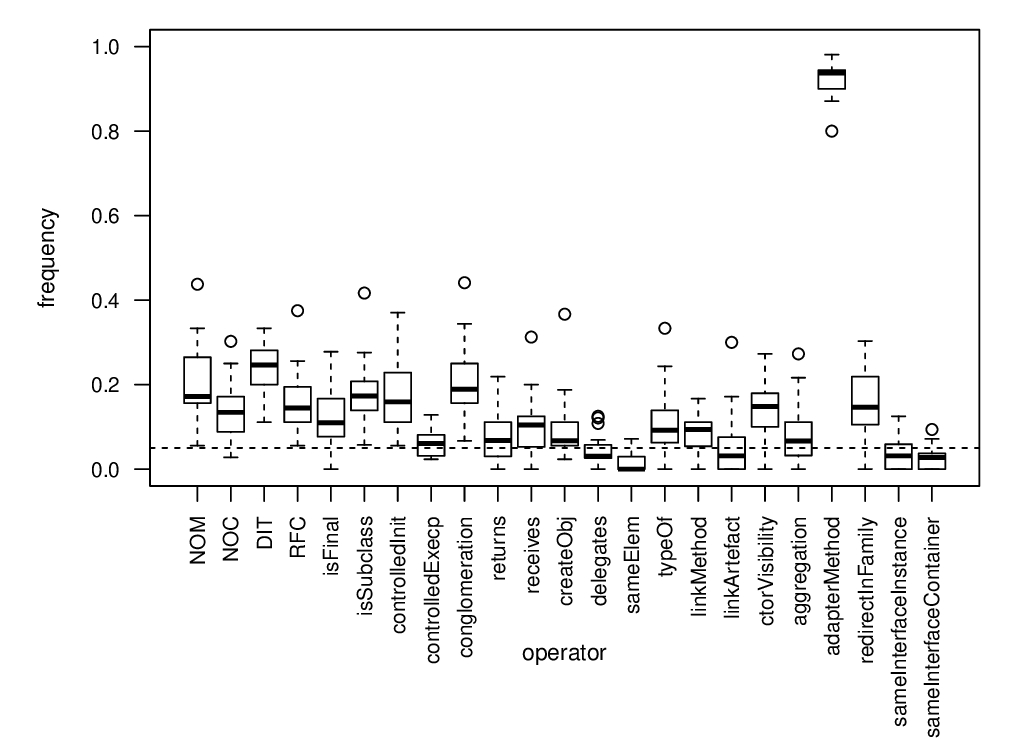}
        \caption{Adapter}
        \label{subfig:norm_adapter-op}
    \end{subfigure}
    \\
    \begin{subfigure}[!t]{0.49\textwidth}
       \includegraphics[scale=0.49]{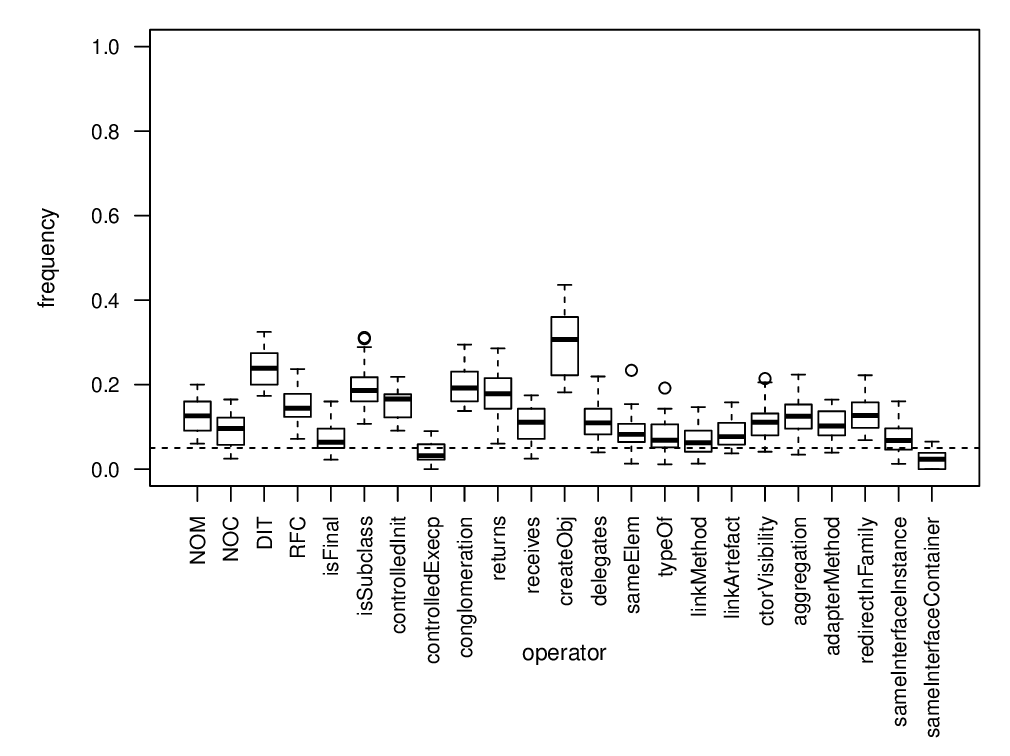}
        \caption{Factory Method}
        \label{subfig:norm_fMethod-op}
    \end{subfigure}
     \hfill
    \begin{subfigure}[!t]{0.49\textwidth}
       \includegraphics[scale=0.49]{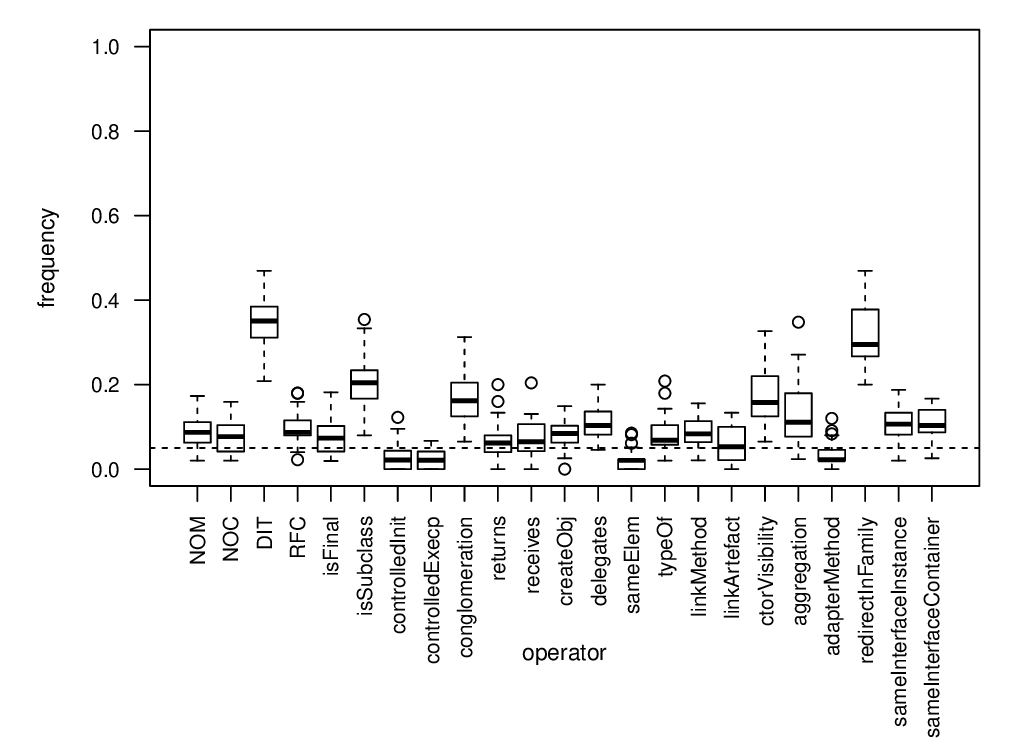}
        \caption{Decorator}
        \label{subfig:norm_decorator-op}
    \end{subfigure}
    \\
    \begin{subfigure}[!t]{0.49\textwidth}
       \includegraphics[scale=0.49]{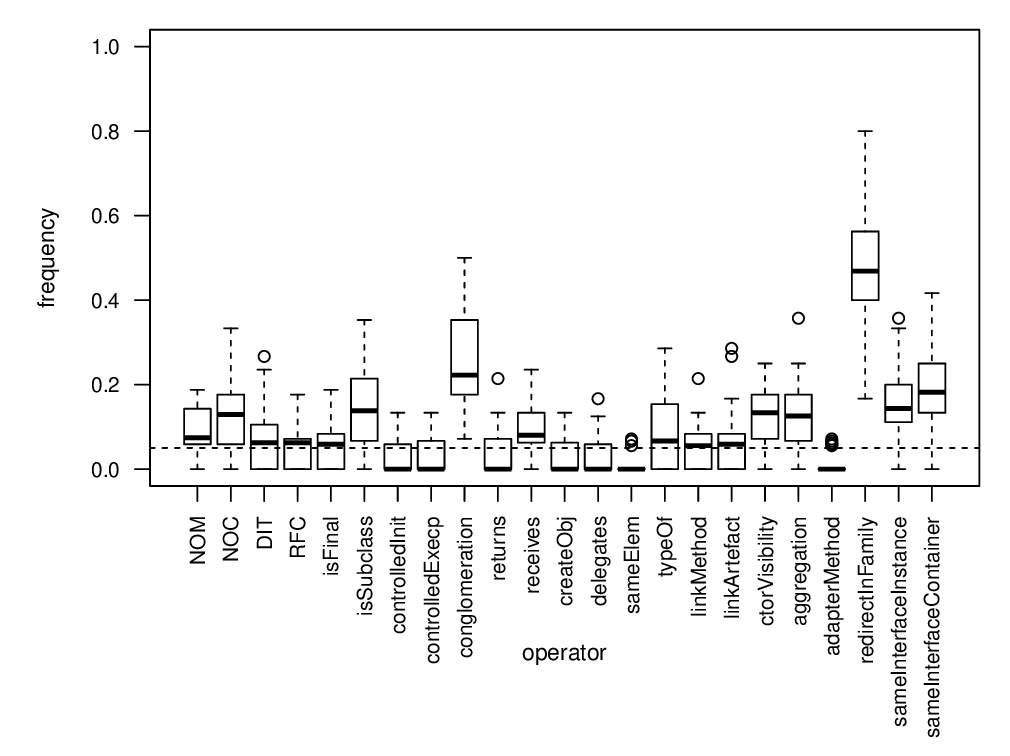}
        \caption{Composite}
        \label{subfig:norm_composite-op}
    \end{subfigure}
    \caption{Frequency of appearance of grammar operators in the resulting rules}
    \label{fig:norm-op}
\end{figure*}

The CFG determines the type of expressions (see Section~\ref{subsec:g3p4dpd_encoding}) to appear within the rules describing DP implementations. As shown in Section~\ref{subsec:exp1_param}, all these operators could be applied without further parametrisation to capture any type of design pattern under analysis, making the practical use of this detection model easier to the software engineer. Nevertheless, engineers could discard from the grammar those operators referred to microstructures that, in their opinion, are not descriptive of the pattern to be detected, or even are modified to comply with their organisational practices. It seems natural to think that the selection of a number of representative operators would produce rules formed from a limited set of pre-selected elements. In addition, it could reduce the search space and, consequently, the time required to find the best rules. Therefore, in response to RQ2, it is interesting to analyse to what extent the selection of operators influence the search process and which operators provide a better detection capability for each design pattern. With this aim, we have counted the number of occurrences in the pruned set of rules for the five design patterns after 30 executions. The results are depicted in Fig.~\ref{fig:norm-op} as box-plots, where the frequency of appearance has been normalised to the range [0, 1].

For the Singleton pattern, it can be observed that those operators related to the class instantiation, such as \textit{ctorVisibility} or \textit{controlledExcep} have a strong presence, whereas those requiring more than one role as input are hardly used. As for the Adapter, \textit{adapterMethod}, which is a pattern-specific operator, appears in most of rules, whereas \textit{sameElem} or \textit{sameInterfaceContainer} are rarely selected. As expected, those operators related to object creation like \textit{createObj} and \textit{returns} commonly appear in the rules describing the samples of the Factory Method. Finally, \textit{redirectInFamily}, \textit{sameInterfaceInstance} and \textit{sameInterfaceContainer} are recurrent operators for the Decorator and Composite patterns, as they are related to delegations between classes belonging to the same inheritance tree. The ability of G3P4DPD to mine negative rules is the reason why almost every operator appears in the final rule set. Notice that using a reduced set of operators would require a smaller number of generations, \ie 100, to complete the search. Besides, including (potentially) dispensable operators within a rule could affect its readability too. Therefore, those operators whose mean frequency is less than or equal to 0.05 have been omitted from the experiments in order to find out how this selection would affect the detection performance. This value, depicted as dashed lines in Fig.~\ref{fig:norm-op}, has been set to all DPs after preliminary experimentation.

The experimentation has been carried out analogously to the previous experiments without selection of grammar operators. For brevity, only the best results will be shown (for a complete list of results, see Additional Material). Table~\ref{tab:resultsOperator} lists the best results obtained for each DP considering the reduced set of grammar operators. Figures in bold typeface represent those values that improve the analogous results --- i.e. same support, confidence and strategy --- obtained without reducing the set of operators (see Table~\ref{tab:resultsAllOperators}). As can be observed, DFML$_{\chi^2}$ is still the classification strategy obtaining the best results for every pattern. Regarding the confidence threshold, best values are mostly reached for configurations with a high value (0.7), the Composite being the only requiring a lower value (0.5). Again, a high support value (0.1) is never recommended. In general, note that the resulting values remain robust and there are not notorious differences with previous experiments including all operators. As for $F_1$, in absolute terms, the selection seemingly benefits the results obtained for Singleton and Composite. In the case of the Adapter, values with or without selecting operators reflect practically a tie. We argue that this consistency in the results favours the engineers feeling able to adapt the selection of grammar operators either to the needs of their organisational repositories or to reduce the search space without significantly affecting the detection performance. Pairwise comparisons have been performed between analogous configurations having and having not reduced the set of grammar operators. This statistical analysis reflects that there are no significant differences, the Factory Method being the only exception, as it slightly improves when all operators are used. 

\begin{table*}[!t]
\caption{Best results for the operator study}
\label{tab:resultsOperator}
\centering
\scriptsize{
\begin{tabular}{c c c c c c c c }
\hline
 & Strategy & Supp-Conf & Accuracy & Precision & Recall & Specificity & $F_1$\\ 
 \hline
 Singleton             & DFML$_{\chi^2}$ & (0.01) - (0.7) & $ \bm{0.9596 \pm 0.0115} $ & $ \bm{0.9492 \pm 0.0180} $ & $ \bm{0.9508 \pm 0.0251} $ & $ \bm{0.9649 \pm 0.0128} $ & $ \bm{0.9452 \pm 0.0173} $ \\
 Adapter               & DFML$_{\chi^2}$ & (0.01) - (0.7) & $ 0.8686 \pm 0.0017 $ & $ 0.8428 \pm 0.0028 $ & $ \bm{0.9125 \pm 0.0026} $ & $ 0.8236 \pm 0.0033 $ & $ 0.8756 \pm 0.0016 $ \\
 Factory Method       & DFML$_{\chi^2}$ & (0.01) - (0.7) & $ 0.8230 \pm 0.0099 $ & $ 0.8097 \pm 0.0082 $ & $ 0.8816 \pm 0.0209 $ & $ 0.7545 \pm 0.0158 $ & $ 0.8423 \pm 0.0107 $ \\
 Decorator            & DFML$_{\chi^2}$ & (0.05) - (0.7) & $ \bm{0.8236 \pm 0.0129} $ & $ \bm{0.8054 \pm 0.0271} $ & $ 0.7196 \pm 0.0246 $ & $ 0.8868 \pm 0.0175 $ & $ 0.7490 \pm 0.0193 $ \\
 Composite            & DFML$_{\chi^2}$ & (0.05) - (0.5) & $ \bm{0.8840 \pm 0.0187} $ & $ \bm{0.7438 \pm 0.0465} $ & $ \bm{0.9089 \pm 0.0564} $ & $ \bm{0.8763 \pm 0.0271} $ & $ \bm{0.7937 \pm 0.0345} $ \\ \hline
\end{tabular}
}
\end{table*}  

To sum up, the following insights about the impact of the grammar operators can be extracted (RQ2):

\begin{itemize}
    \item \ourtechnique is able to automatically determine the type of operators more relevant to each category of design pattern. Those related to visibility and instantiation allow detecting creational patterns (Singleton and Factory method), whereas operators focused on delegation structures are highly effective to recover structural and behavioural patterns (Adapter, Decorator and Composite).
    \item The general performance of \ourtechnique does not significantly decrease when the set of operators is reduced, Factory method being the only exception.
\end{itemize}

\subsection{Performance Comparison and Discussion}\label{subsec:exp1_discussion}

\paragraph{Performance comparison} Section~\ref{subsec:exp1_operators} showed how robust \ourtechnique behaves when one specific part of the configuration, \ie the grammar operators representing design microstructures, is refined. Nevertheless, even though it is likely that software engineers could make design-based decisions like selecting these microstructures, it still seems unrealistic to think that they would have the skills required to set up the rest of parameters and adjust the model to their needs. In these sense, for the sake of practicality, we have considered the use of a common model configuration for all design patterns taking the results listed in Section~\ref{subsec:exp1_param}. As previously observed, DFML$_{\chi^2}$ clearly dominates the rest of classification strategies. As for the support and confidence thresholds, S(0.01)-C(0.7) reaches the best values for the Singleton and Adapter, whereas S(0.05)-C(0.7) is the most appropriate for the Factory Method and Decorator. Finally, S(0.05)-C(0.6) was the best configuration for the Composite pattern. As lower values of support are preferred in order to find rare DP implementations, S(0.01) is seemingly a comprehensive choice. Similarly, C(0.7) is selected as the confidence threshold since it reaches the best values for all the cases, except for the Composite, for which also gets competitive performance values. The results obtained for this configuration for the five patterns are also shown in Table~\ref{tab:comparison}, which compiles the results obtained during the parameter study. More precisely, it shows the accuracy and $F_1$ returned by the best configuration ---in terms of $F_1$--- found for each design pattern, as well as the default common configuration discussed above. 

\begin{table*}[!t]
\caption{Comparison results for Experiment \#1}
\label{tab:comparison}
\centering
\small{
\begin{tabular}{c c c c c | c c c c}
\hline
& \multicolumn{4}{c}{\raggedright \textbf{Best configuration}} & \multicolumn{4}{c}{\raggedright \textbf{General-purpose configuration}}  \\ \hline
& \multicolumn{2}{c}{\raggedright \textbf{\ourtechnique}} & \multicolumn{2}{c}{\raggedright \textbf{MARPLE}} &
\multicolumn{2}{c}{\raggedright \textbf{\ourtechnique}} & \multicolumn{2}{c}{\raggedright \textbf{MARPLE}}  \\
\hline
 & Accuracy & $F_1$ & Accuracy & $F_1$ & Accuracy & $F_1$ & Accuracy & $F_1$\\ 
 \hline
 Singleton          & \textbf{0.9561} & \textbf{0.9411} & 0.88 & 0.91 & \textbf{0.9561} & \textbf{0.9411} & 0.93 & 0.90 \\ 
 Adapter            & \textbf{0.8688} & \textbf{0.8757} & 0.86 & 0.85 & \textbf{0.8688} & \textbf{0.8757} & 0.85 & 0.84 \\ 
 Factory Method     & \textbf{0.8304} & \textbf{0.8503} & 0.82 & 0.83 & \textbf{0.8308} & \textbf{0.8489} & 0.82 & 0.83 \\
 Decorator          & 0.8229 & 0.7501 & 0.82 & \textbf{0.77} & 0.8188 & 0.7448 & 0.82 & \textbf{0.77} \\
 Composite          & \textbf{0.8859} & \textbf{0.7885} & 0.77 & 0.56 & \textbf{0.8888} & \textbf{0.7568} & 0.75 & 0.45 \\ \hline
\end{tabular}
}
\end{table*}

In response to RQ3, the detection performance of \ourtechnique is compared against MARPLE, a ML-based tool experimented under the same experimental methodology. More specifically, we compare the best results obtained by MARPLE for each pattern, as originally reported by the authors~\citep{zanoni2015}. The general-purpose configuration for MARPLE corresponds to a Random Forest classifier with k-Means as preprocessor, as this combination provided the best results for three out of the five patterns. Comparative results can be found in Table~\ref{tab:comparison}, considering the best and the general-purpose configuration for both techniques. As can be observed, considering the five DPs in this experiment, \ourtechnique outperforms MARPLE for the Singleton, Adapter and Factory Method, with percentage of improvements equal to 4.57\%, 4.25\% and 2.28\%, respectively, when comparing general-purpose configurations. In contrast, MARPLE achieves an increase of 2.65\% in $F_1$ for the Decorator. However, \ourtechnique demonstrates a significant advantage in the case of the Composite pattern, obtaining an increase of 35.14\% in $F_1$ when comparing the results of our general-purpose configuration even against the best results of MARPLE, and 68.18\% if we compare their best configurations. Again, it is worth remarking that \ourtechnique reaches more stable values along the different design patterns, in contrast to MARPLE.

\ourtechnique provides better detection performance for the Singleton pattern, whose structure exhibits one single role. In contrast, the worst results are returned for the Decorator and Composite. This is in line with the conclusions drawn by other DPD models~\citep{uchiyama2011,chihada2015,zanoni2015}, where the detection capacity decreases as the pattern complexity increases. 

From the comparison results, the following findings related to RQ3 can be extracted:

\begin{itemize}
    \item \ourtechnique outperforms MARPLE for four out the five patterns under comparison, achieving up to 68.18\% of improvement in terms of F1 for the Composite pattern.
    \item \ourtechnique provides more stable results than MARPLE, with more than 81\% of accuracy and 74\% of F1 for all patterns, using either its general-purpose or best configuration.
    \item Like other ML proposals, \ourtechnique shows a better detection performance for less complex patterns.
\end{itemize}

\paragraph{Discussion of design characteristics} \ourtechnique presents some characteristics in its design that makes it an appealing alternative beyond its good performance. We next discuss their implications from a more practical perspective. The use of a CFG and a large, diverse set of eligible design properties in terms of grammar operators facilitates the customisation of the detection process to the engineers' needs. In contrast to other proposals, \ourtechnique allows both numerical and categorical properties, meaning the simultaneous use of metrics and design microstructures. In this way the software engineer can explicitly influence the design and developmental aspects that will guide the search. Furthermore, the resulting model would be able to capture the most significant properties in each moment, adapting the model to changes in the repository. Promoted by the use of ML techniques, new detected samples --- both positive and negative --- by the resulting models could also serve as an input for future detections. Thus, \ourtechnique would allow the progressive adjustment of the prediction to the organisational development culture, what is usually a dynamic element in companies. Finally, in comparison to other black-box proposals~\citep{chihada2015,Dwivedi2018,thaller2019}, \ourtechnique uses a rule-based model, which is more comprehensible for practitioners~\citep{Kotsiantis06}.


\section{\new{Experiment 2: Comparison Against DPD Methods Using P-Mart}}
\label{sec:experiment2}

\begin{table*}[!t]
\caption{\new{Comparison results for JHotDraw project.}}
\label{tab:jhotdrawResults}
\centering
\scalebox{0.68}{
\begin{tabular}{cc|ccccc|ccccc|ccccc|ccccc}
\hline
& Ground &\multicolumn{5}{c|}{\raggedright \ourtechnique} & \multicolumn{5}{c|}{\raggedright DePATOS} & \multicolumn{5}{c|}{\raggedright SparT} & \multicolumn{5}{c}{\raggedright MLDA} \\
& truth & P & TP & Pr & Re & $F_1$ & P & TP & Pr & Re & $F_1$ & P & TP & Pr & Re & $F_1$ & P & TP & Pr & Re & $F_1$\\
 \hline
 Adapter            & 1 & 1 & 1 & 1.00 & 1.00 & 1.00 & 38 & 1 & 0.03 & 1.00 & 0.05 & 12 & 1 & 0.08 & 1.00 & 0.15 &        19 & 1 & 0.05 & 1.00 & 0.10\\ 
 Command            & 1 & 11 & 1 & 0.09 & 1.00 & 0.17 & - & - & - & - & - & 8 & 1 & 0.13 & 1.00 & 0.22 &                 12 & 1 & 0.08 & 1.00 & 0.15\\ 
 Composite          & 1 & 11 & 1 & 0.09 & 1.00 & 0.17 & 1 & 1 & 1.00 & 1.00 & 1.00 & 1 & 1 & 1.00 & 1.00 & 1.00 &        1 & 1 & 1.00 & 1.00 & 1.00\\ 
 Decorator          & 1 & 2 & 1 & 0.50 & 1.00 & 0.67 & 3 & 0 & 0.00 & 0.00 & 0.00 & 3 & 1 & 0.33 & 1.00 & 0.50 &         1 & 1 & 1.00 & 1.00 & 1.00\\ 
 Factory Method     & 2 & 0 & 0 & 0.00 & 0.00 & 0.00 & - & - & - & - & - & 2 & 1 & 0.50 & 0.50 & 0.50 &                  0 & 0 & 0.00 & 0.00 & 0.00\\ 
 Observer           & 2 & 26 & 2 & 0.08 & 1.00 & 0.14 & - & - & - & - & - & 4 & 0 & 0.00 & 0.00 & 0.00 &                 0 & 0 & 0.00 & 0.00 & 0.00\\
 Singleton          & 2 & 2 & 2 & 1.00 & 1.00 & 1.00 & - & - & - & - & - & 2 & 2 & 1.00 & 1.00 & 1.00 &                  2 & 2 & 1.00 & 1.00 & 1.00\\ 
 State/Strategy     & 6 & 7 & 1 & 0.14 & 0.17 & 0.15 & - & - & - & - & - & 24 & 2 & 0.08 & 0.33 & 0.13 &                 22 & 3 & 0.13 & 0.50 & 0.20\\ 
 Template Method    & 2 & 4 & 2 & 0.50 & 1.00 & 0.67 & - & - & - & - & - & 5 & 1 & 0.20 & 0.50 & 0.29 &                  - & - & - & - & -\\ 
 \hline
 \textbf{Total}     & 18 & 64 & 11 &  &  &  & 42 & 2 &  &  &  & 61 & 10 & & & & 57 & 9 & &\\
 \hline
\end{tabular}
}
\end{table*}

\new{This section presents the results and analysis of Experiment \#2, in which GEML is compared to other DPD studies considering a project available in P-Mart, JHotDraw, as the testing project. We provide results from 10 out of the 11 Gamma’s DPs available in this project, since the rest of P-Mart projects cannot provide training instances for the Prototype pattern. Three recent DPD methods are chosen for comparison: DePATOS, which detects structural patterns using a sub-graph isomorphism algorithm~\citep{yu2018}; MLDA, a rule-based approach that analyses method signatures~\cite{Al-Obeidallah2019}; and SparT, a method based on ontologies that combines structural, behavioural and semantic information~\cite{Xiong2019}. All these works provide results on JHotDraw that can be contrasted, since their authors both report absolute values of recovered instances, and give access to the DP implementations found. This allows us to compute the performance metrics (precision, recall, and F1) on the basis of a common ground truth, instead of comparing results validated over custom repositories created by the authors. Notice that, usually, P-Mart is extended in DPD studies, but the resulting datasets are not publicly available.}

\new{Table~\ref{tab:jhotdrawResults} lists the DPs under study and the results of the four methods in terms of number of recovered instances per DP, as well as the total sum. Precision, recall and F1 are computed considering positive samples labelled in P-Mart as the ground truth for all methods. None of the methods under comparison distinguish State from Strategy, so we report them together for GEML\footnote{Notice that GEML can detect both State and Strategy design patterns separately.} too. The symbol ``-'' is used to indicate that the particular DP is not supported by the method. GEML is the method that recovers more true DP implementations (11), followed by SparT (10) and MLDA (9). In terms of F1, GEML is the best method for four DPs (Adapter, Observer, Singleton and Template method), and the second best method for the remaining DPs. The only DP for which no implementation is found is the Factory method, which is not supported by DePATOS and not detected by MLDA. GEML considerably reduces the number of false positives for the Adapter and State/Strategy patterns compared to the rest of methods. In contrast, given that GEML is a ML-based approach, it suffers from the low number of training instances in some cases, such as Composite and Observer. Even so, GEML manages to detect the same number of true positives than the other methods for the Composite pattern, and it is the only method that is able to recover the two Observer implementations.}

\section{Experiment \new{3}: Analysis of Applicability}
\label{sec:experiment3}

Experiment \#\new{3} is explained in this section on the basis of a practical scenario, and evaluated according to more qualitative aspects. The experimentation has been performed with a large number of DPs (15), showing that no adaptation is required to execute \ourtechnique when new DPs are introduced. The outcomes serve us to analyse how the change of the training repository might influence the behaviour of the proposed method (RQ4). Then, a comparison against reference non-ML-based DPD tools is provided (RQ5).

\subsection{Influence of Training Factors}\label{subsec:exp2_performance}

In response to RQ4, \ourtechnique is retrained by applying its general-purpose configuration and using a different input repository than in Experiment \#1, as explained in Section~\ref{subsec:methodology}. Then, the detection model is tested on DPExample. Table~\ref{tab:exp2_radd_results} shows the results for the median and best execution of the 30 runs. \textit{Pr} and \textit{Re} stand for precision and recall, respectively. Design patterns are sorted in decreasing number of positive training samples (see Table~\ref{tab:candidates}).

\begin{table*}[!t]
\caption{Performance of \ourtechnique in the test project (DPExample) with and without using numerical properties}
\label{tab:exp2_radd_results}
\centering
\small{
\begin{tabular}{c ccc| ccc | ccc | ccc}
\hline
& \multicolumn{6}{c}{\raggedright \textbf{With numerical properties}} & \multicolumn{6}{c}{\raggedright \textbf{Without numerical properties}}  \\ \hline
& \multicolumn{3}{c}{\raggedright Median} & \multicolumn{3}{c}{\raggedright Best} &
\multicolumn{3}{c}{\raggedright Median} & \multicolumn{3}{c}{\raggedright Best}  \\
\hline
 & Pr & Re & $F_1$ & Pr & Re & $F_1$ & Pr & Re & $F_1$ & Pr & Re & $F_1$\\ 
 \hline
 State              & 0.09 & 0.67 & 0.16 & 0.13 & 0.67 & 0.22 & 0.10 & 0.67 & 0.18 & 0.14 & 0.67 & \textbf{0.23} \\ 
 Adapter            & 0.04 & 0.58 & 0.08 & 0.12 & 0.83 & 0.21 & 0.04 & 0.67 & 0.08 & 0.17 & 0.33 & \textbf{0.22} \\ 
 Singleton          & 0.81 & 0.81 & 0.81 & 0.81 & 0.81 & \textbf{0.81} & 0.81 & 0.81 & 0.81 & 0.81 & 0.81 & \textbf{0.81} \\ 
 Template Method    & 0.22 & 0.21 & 0.20 & 0.55 & 0.86 & \textbf{0.67} & 0.40 & 0.86 & 0.55 & 0.55 & 0.86 & \textbf{0.67} \\ 
 Proxy              & 0.38 & 0.75 & 0.50 & 0.50 & 0.75 & 0.60 & 0.43 & 0.75 & 0.55 & 0.60 & 0.75 & \textbf{0.67} \\
 Observer           & 0.14 & 0.29 & 0.19 & 0.67 & 0.29 & 0.40 & 0.50 & 0.14 & 0.22 & 1.00 & 0.29 & \textbf{0.44} \\
 Strategy           & 0.09 & 0.50 & 0.15 & 0.10 & 0.50 & 0.16 & 0.09 & 0.50 & 0.15 & 0.11 & 0.63 & \textbf{0.19} \\
 Composite          & 0.10 & 0.50 & 0.17 & 0.12 & 0.50 & 0.19 & 0.10 & 0.50 & 0.17 & 0.13 & 0.67 & \textbf{0.22} \\
 Abstract Factory   & - & - & - & 0.04 & 0.37 & 0.08 & - & - & - & 0.16 & 0.79 & \textbf{0.25} \\
 Command            & - & - & - & 0.01 & 0.40 & 0.02 & 0.01 & 0.40 & 0.02 & 0.01 & 0.20 & \textbf{0.02} \\
 Factory Method     & - & - & - & - & - & - & - & - & - & 0.12 & 0.20 & \textbf{0.15} \\
 Visitor            & 0.36 & 0.93 & 0.52 & 0.42 & 0.93 & \textbf{0.58} & 0.36 & 0.93 & 0.52 & 0.42 & 0.93 & \textbf{0.58} \\
 Iterator           & 0.02 & 0.40 & 0.03 & 0.80 & 0.80 & \textbf{0.80} & 0.02 & 0.40 & 0.03 & 0.67 & 0.80 & 0.73 \\
 Decorator          & 0.75 & 0.33 & 0.45 & 0.80 & 0.67 & \textbf{0.73} & 1.00 & 0.33 & 0.50 & 0.80 & 0.67 & \textbf{0.73} \\
 Bridge             & 0.04 & 0.17 & 0.07 & 0.07 & 0.50 & \textbf{0.13} & 0.04 & 0.17 & 0.07 & 0.07 & 0.50 & \textbf{0.13} \\
 \hline
\end{tabular}
}
\end{table*}

The results bring new insights. Despite the low number of training instances in the repository ---a limiting factor inherent to machine learning--- \ourtechnique is able to retrieve DP instances for 15 design patterns, all Gamma's DP available in P-Mart, with the exception of Builder, Memento, Prototype and Facade. A general observation is that our method suffers when less than 10 training samples are provided. In these cases, it might happen that the lack of data result in executions for which no detection rules can be generated (marked as '-' in Table~\ref{tab:exp2_radd_results}). The degradation in performance is more evident in terms of precision, \ie the rate of false positives tends to increase. Even so, \ourtechnique finds at least half of the true DP implementations during its best execution, with the exception of four design patterns (Observer, Abstract Factory, Command and Factory Method).

Another consequence of having a reduced training set is overfitting, since less variability of examples is presented to the algorithm. The fact that DPExample is the only project taken from the literature, in contrast to other open source projects used for training, may cause this project to present slight differences for some properties relying on role identification and software metrics. This is specially evident for two DPs, Adapter and Template Method, since \ourtechnique provides a considerably higher precision and recall in the training phase. For the Adapter, we have observed differences in how DPB and P-Mart label the roles, since DPB assigns one class per role at most~\citep{zanoni2015}. This affects the effectiveness of the \textit{adapterMethod} operator, which works under the same assumption. Similarly, we realised that numerical properties (software metrics) become less informative in this experiment. The difference in size between classes inspected during training and those available for testing implies that the learned thresholds for measures like LOC and NOM are not so representative for the testing project. After excluding such operators, our method has improved its detection capability for six design patterns and has guaranteed more stable results for Abstract Factory, Command and Factory Method (see Table~\ref{tab:exp2_radd_results}). The generated rule set for each DP is available as additional material.

The specificity of some operators has revealed as an important factor to counteract the low number of training samples. This is reflected in the Decorator pattern, for which \ourtechnique provides similar results than those obtained in Experiment \#1 after learning from two positive samples only in Experiment \#2. Looking at the resulting detection models, we observe that \textit{redirectInFamily} --- an operator particularly associated to the Decorator pattern --- frequently appears in the rules, as shown in Section~\ref{subsec:exp1_operators}. This finding does not imply that new operators oriented towards a particular DP have to be implemented to support its detection. This can be observed in the case of the Visitor pattern, which has obtained the highest recall despite being a pattern with very few positive samples and not being considered in Experiment \#1. For this specific pattern, the operator checking whether a role is implemented by an interface becomes highly relevant to identify its structure. Finally, State and Strategy share the same class structure, making it difficult to distinguish them and therefore to detect them correctly. For this reason, some DPD methods consider them together. According to our outcomes, \ourtechnique is able to differentiate State from Strategy, and vice versa, for around 10\% of instances, and the corresponding false positives are often attributed to an usual misclassification between both DPs.

Finally, Table~\ref{tab:exp2_time} provides the average execution time of each step of the DPD process applied to DPExample, a medium-size project. G3P4DPD is the step requiring more time, although it does not usually takes longer than one minute. Pruning the generated rules is a very fast procedure that mostly depends on the number of rules returned by G3P4DPD. Note that training the DPD model, \ie rule generation and pruning, only needs to be carried out when new DP instances are added to the repository. The VF2 algorithm is efficient for candidate generation, and no great disparity is observed for design patterns with different number of roles. Then, the filtering step takes two seconds at most, depending on the different heuristics applied to each DP (see Section~\ref{subsec:methodology}). Finally, the time required to proceed with the detection depends on both the number of candidates and the number of rules.

\begin{table}[!t]
\caption{Average execution time (ms) of each phase of the DPD process. Graph building from code takes 909.67 ms on average.}
\label{tab:exp2_time}
\centering
\scriptsize{
\begin{tabular}{c|cc|cc|c}
\hline
& \multicolumn{2}{c|}{Rule} & \multicolumn{2}{c|}{Candidates} & DP \\
& gener. & pruning & gener. & filtering & detection \\
\hline
 State              & 68,188.23 & 29.27 & 24.00 & 168.37 & 1,066.80\\     
 Adapter            & 27,949.63 & 19.20 & 22.37 & 300.87 & 722.67 \\ 
 Singleton          & 19,112.73 & 6.50  & 15.57 & 204.23 & 1,140.40 \\    
 T. Method          & 39,412.37 & 17.83 & 18.20 & 178.30 & 584.30 \\    
 Proxy              & 12,099.03 & 3.33  & 18.47 & 160.83 & 85.83 \\     
 Observer           & 5,157.50 & 67.67  & 24.50 & 75.37 & 60.43 \\     
 Strategy           & 6,980.83 & 1.63   & 24.03 & 260.57 & 197.23  \\     
 Composite          & 14,909.90 & 1.40  & 22.30 & 237.47 & 697.50 \\     
 A. Factory         & 8,852.90 & 1.20   & 23.20 & 1,733.53 & 52.67 \\     
 Command            & 11,596.50 & 0.97  & 37.30 & 240.57 & 342.03 \\     
 F. Method          & 5,145.03 & 0.80   & 31.20 & 160.07 & 2.67 \\     
 Visitor            & 9,765.87 & 0.70   & 17.13 & 27.93 & 3.13 \\ 
 Iterator           & 3,928.11 & 0.74   & 23.95 & 14.03 & 110.71 \\
 Decorator          & 4,174.20 & 0.50   & 35.43 & 434.97 & 3.93 \\
 Bridge             & 3,961.95 & 0.72   & 31.55 & 158.61 & 4.10 \\
 \hline
 
\end{tabular}
}
\end{table}

In light of these results, the influence of the training conditions (RQ4) can be summarised as follows:

\begin{itemize}
    \item The detection performance of \ourtechnique reaches up to 81\%, though it varies depending on the design pattern. \ourtechnique achieves more than 50\% of recall for 11 out of the 15 DPs, and more than 75\% for six of them, despite the low number of samples available.
    \item The detection capability is not only affected by the training set size, but also the particular characteristics of each DP and how roles are labelled in the repository.
    \item The choice of operators, especially not using software metrics, become more relevant when few samples are available. The improvement can be up to 300\% in terms of precision, 114\% in terms of recall and 213\% with respect to F1.
    \item \ourtechnique is able to learn rules for any new design pattern without requiring the implementation of specific operators, but might have difficulties to produce rules for some DPs when the number of samples is significantly low.
\end{itemize}

\subsection{Comparison with DPD Tools}\label{subsec:exp2_tools}

In the context of RQ3, the previous results should be contrasted with those obtained with other tools available to the software engineer, \ie SSA and Ptidej. It should be noted that the conditions under which this comparison can be carried out (see Section~\ref{subsec:methodology}) are not favourable for \ourtechnique. On the one hand, the number of training samples for some DPs is extremely low. On the other hand, since \ourtechnique has partially learned from the outcomes of both tools, \ourtechnique might fail to identify the instances not discovered by these tools. Even so, \ourtechnique manages to be superior, or at least competitive, to SSA and Ptidej. Table~\ref{tab:exp2_comparison} shows the best results for \ourtechnique together with those obtained after running both tools. We report absolute numbers of positive DP implementations retrieved (P) and correctly identified (TP) by each method, since they seem easier to interpret and give an idea of the effort required to manually verify tool outcomes. As a reference, the actual number of DP implementations (ground truth) is provided too.

\begin{table}[!t]
\caption{Comparison results for Experiment \#2}
\label{tab:exp2_comparison}
\centering
\scriptsize{
\begin{tabular}{cc|cc|cc|cc}
\hline
& Ground &\multicolumn{2}{c|}{\raggedright \tiny{\ourtechnique}} & \multicolumn{2}{c|}{\raggedright SSA} & \multicolumn{2}{c}{\raggedright Ptidej} \\
& truth & P & TP & P & TP & P & TP \\
 \hline
 State              & 12    & 57 & 8  & 41 & 3    & 104 & 9\\     
 Adapter            & 6     & 12 & 2  & 54 & 3    & 128 & 3\\ 
 Singleton          & 21    & 21 & 17 & 22 & 17   & 82 & 15\\    
 T. Method          & 7     & 11 & 6  & 20 & 7    & 234 & 7\\    
 Proxy              & 4     & 5 & 3   & 2 & 2     & 127 & 4\\     
 Observer           & 7     & 2 & 2   & 4 & 2     & - & -  \\     
 Strategy           & 8     & 46 & 5  & 6 & 6     & 0 & 0  \\     
 Composite          & 6     & 30 & 2  & 7 & 1     & 29 & 2 \\     
 A. Factory         & 19    & 99 & 15 & - & -     & - & -  \\     
 Command            & 5     & 79 & 1  & 4 & 3     & 36 & 2 \\     
 F. Method          & 15    & 26 & 3  & 1 & 1     & 44 & 5 \\     
 Visitor            & 15    & 33 & 14 & 11 & 11   & 2 & 2  \\    
 Iterator           & 5     & 5  & 4  & - & -    & - & - \\
 Decorator          & 6     & 5 & 4   & 19 & 5    & - & -  \\
 Bridge             & 6     & 41 & 3 & 3 & 0 & - & - \\
 \hline
 \textbf{Total}     & 142   & 472  & 89  & 194 & 61 & 786 & 49\\
 \hline
\end{tabular}
}
\end{table}

Overall, \ourtechnique detects 63\% of the DP implementations, followed by SSA (43\%) and Ptidej (35\%). SSA applies a more conservative detection strategy that allows reducing the presence of false positives (FPs). Proxy, Strategy, Factory Method, Bridge and Visitor clearly illustrate this point. \ourtechnique shows more variability in this regard, with more than a half of FPs corresponding to three DPs: Abstract Factory (not supported by the other tools), Command and State. The issues discussed in previous Section~\ref{subsec:exp2_performance}, such as the low number of instances and overfitting, are the reason behind such behaviour. Even so, \ourtechnique considerably reduces the need of inspecting all classes within the project, and tends to return less FPs than Ptidej. We observe that this tool correctly identifies the classes implementing the design pattern, but fails to assign the roles and returns several permutations as solution for the same DP instance. SSA presents some limitations with respect to role identification too. For 12 DPs, SSA does not provide the classes playing one or more roles, meaning that the practitioner is required to manually inspect the code to complete the DP definition.

Each tool appears to be superior for a different set of design patterns, not necessarily those belonging to the same category (creational, structural or behavioural). Indeed, the only DPs for which the three tools return the majority of implementations are Singleton and Template Method. Difficulties to detect implementations of Factory Method, Composite and Adapter are observed in the three tools. SSA finds one DP instance more than \ourtechnique for Adapter, Template Method, Strategy and Decorator, and two more instances of the Command pattern. Higher differences are observed in favour of \ourtechnique for State (5), Bridge (3), Visitor (3) and Factory Method (2). Compared to Ptidej, \ourtechnique also returns a similar number of correct DP instances for State, Adapter, Template Method and Proxy. However, the lack of support to four DPs (Observer, Bridge, Abstract Factory and Decorator) and the fact that Strategy and State are considered together impose some limitations to Ptidej.

\begin{table*}[!t]
\caption{Coincidences among DPD tools}
\label{tab:exp2_coincidences}
\centering
\small{
\begin{tabular}{cc|cccc|cccc}
\hline
& & \multicolumn{4}{c|}{Coincidences in positive implementations} & \multicolumn{4}{c}{Coincidences in true positive implementations}\\\cline{3-10}
& Unique & \scriptsize{\ourtechnique} & \scriptsize{\ourtechnique} & SSA $\cap$ & All & \scriptsize{\ourtechnique} & \scriptsize{\ourtechnique} & SSA $\cap$ & All \\
& TP     & $\cap$ SSA    & $\cap$ Ptidej & Ptidej     &  & $\cap$ SSA    & $\cap$ Ptidej & Ptidej     & \\
 \hline
 State           & 1  &38.0\% &53.3\% &33.0\% & 24.6\%   & 37.5\% & 70.0\%& 33.3\%& 30.0\%\\  
 Adapter         & 0  &15.8\% &6.8\% &20.6\% & 4.2\%     & 75.0\% & 75.0\%& 100.0\%& 75.0\%\\ 
 Singleton       & 0  &95.4\% &17.15\% &16.9\% & 16.9\%  & 100.0\% & 68.4\% & 68.4\% & 68.4\%\\ 
 T. Method       & 0  &52.4\% &5.1\% &8.6\% & 4.7\%      & 85.7\% & 85.7\%& 100.0\%& 85.7\%\\  
 Proxy           & 0  &40.0\% &1.5\% &0.8\% & 0.8\%      & 66.7\% & 50.0\%& 25.0\%& 25.0\%\\ 
 Observer        & 2  &0.0\% &- &- & -                   & 0.0\% & - & - & - \\ 
 Strategy        & 0  &10.6\% &- &- & -                  & 83.3\% & - & - & -\\ 
 Composite       & 2  &23.3\% &25.5\% &12.5\% & 8.5\%    & 25.0\% & 50.0\% & 50.0\% & 25.0\%\\ 
 A. Factory      & 15 &- &- & - &-                       & 0.0\% & - & - & - \\
 Command         & 0  &2.4\% &3.6\% &8.6\% & 0.0\%       & 40.0\% & 0.0\% & 75.0\% & 0.0\%\\ 
 F. Method       & 0  &0.0\% &40.0\% &0.0\% & 0.0\%      & 0.0\% & 60.0\% & - & 0.0\%\\ 
 Visitor         & 3  &78.8\% &6.1\% &18.2\% & 6.1\%     & 83.9\% & 6.5\%& 18.2\% & 6.5\% \\ 
 Iterator        & 4  & - & - & - & -                    & - & - & - & - \\
 Decorator       & 1  &14.3\% &- &- &-                   & 50.0\% & - & - & -\\ 
 Bridge          & 3 & 7.3\% & - & - & -                 & 0.0\% & - & - & - \\
 \hline
\end{tabular}
}
\end{table*}

At this point the level of detection agreement between tools is discussed based on the intersection of their result sets so as not to focus solely on the number of DP instances detected. These coincidences are expressed as percentages in Table~\ref{tab:exp2_coincidences}, what includes outcomes for both positive DP implementations and TPs. In light of the results, it does not seem evident to conclude which pair of tools provide more similar results. Due to the FP rate of Ptidej, \ourtechnique is closer to SSA in terms of positive implementations. Nevertheless, \ourtechnique achieves higher agreement with Ptidej for State and Proxy compared to the coincidences between SSA and Ptidej.

Next, we focus on those design patterns for which tools exhibit less agreement. \ourtechnique and SSA both found two out of the seven implementations of the Observer, but they returned different instances. Similarly, the implementation found by SSA for the Factory Method was not detected by \ourtechnique. This phenomenon is observed for Proxy implementations too, for which no pair of tools returns more than two equal instances. Furthermore, Table~\ref{tab:exp2_coincidences} also provides the number of TPs that only \ourtechnique was able to find. In total, \ourtechnique detects 12 implementations missed by SSA or Ptidej, not counting the Abstract Factory nor the Iterator instances, since neither SSA nor Ptidej consider these DPs. For the Observer, the Bridge and the Composite patterns, these DP implementations are the only TPs returned by \ourtechnique, meaning that it is highly effective to find instances that other tools would miss. As for the Visitor, apart from the three unique implementations returned, \ourtechnique is able to find all the instances (11) detected by SSA. This is not what happens more often and, even though each tool has its own ability to find out certain DP implementations, in most cases their results are complementary.

To conclude, we next compile the most relevant facts that give response to RQ5:

\begin{itemize}
    \item \ourtechnique finds more true positive instances than the two reference tools, SSA and Ptidej, including samples of the Abstract Factory pattern and Iterator (not available in these tools).
    \item Considering only those DPs supported by all tools, 11\% of the samples recovered by \ourtechnique were not found by any other tool used for comparison.
    \item The level of agreement among all tools, in terms of true positive implementations, can reach more than 80\%, but for some DPs is significantly lower. Tools for DPD are mutually complementary in terms of practical use.
\end{itemize}

\section{Demonstration Tool}\label{sec:tool}

\ourtechnique is publicly available as a Java-based demonstration tool (see Additional Material) that allows engineers to detect DP implementations from their own projects without requiring any expertise in ML or evolutionary techniques. The tool provides basic graphical support for the whole DPD process, divided into the following three phases:

\begin{enumerate}
\item \textit{Generation of candidates}. The source code is analysed to extract an initial set of potential DP implementations (candidates). 
\item \textit{Learning of the detection model}. The G3P4DPD and pruning algorithms are executed to generate the set of detection rules.
\item \textit{Recovery of design patterns}. The detection rules, together with a classification strategy chosen by the software engineer, check whether the candidates are actually implementing a DP.
\end{enumerate}

\begin{figure}[!t]
\centerline{\includegraphics[scale=0.83]{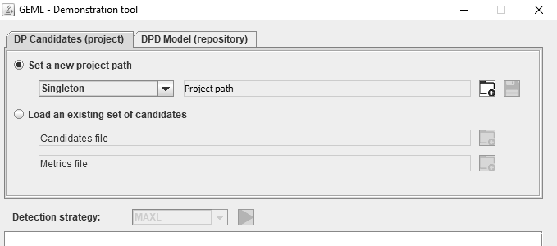}}
\caption{Step 1: Generation of candidates}
\label{fig:dpdtool-candidates}
\end{figure}

Each of these phases is detailed next. Firstly, for the \textit{generation of candidates}, the project from which the set of potential samples for a given DP will be extracted must be selected. During this process, code artefacts and their relationships are scrutinised to find groups of related elements. Then, a role mapping procedure assigns a role to every artefact comprising the candidate according to its relationships. Taking the Adapter pattern as an example, the code artefact playing the \textit{adapter} role has to implement the interface declared by \textit{target}, while adapting the service provided by \textit{adaptee}. Thus, it should be linked to at least two other code artefacts. Similarly, in the case of the Singleton pattern, each code artefact would correspond to a DP candidate, as it only has one role. Fig.~\ref{fig:dpdtool-candidates} shows the screenshot with this configuration panel. Once candidates are found, they can be exported for future executions. A file containing software metrics is also generated. 

\begin{figure}[!t]
\centerline{\includegraphics[scale=0.83]{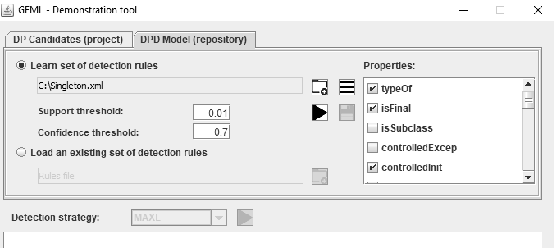}}
\caption{Step 2: Learning of the detection model}
\label{fig:dpdtool-model}
\end{figure}

In a different view, the \textit{learning of the detection model} is conducted, the support and confidence thresholds being set as shown in Fig.~\ref{fig:dpdtool-model}. The user could also select the set of operators to be applied for mining the rules (see Sections~\ref{subsubsec:g3p4dpd_operators} and~\ref{subsec:exp1_operators}). The rest of parameters are set to their default values for simplicity, though advanced users could still modify them by simply editing an XML configuration file. This file also includes the path of the repository from which rules will be mined, so the user could modify it according to organisational or team requirements. Additionally, the resulting detection rules can be saved for future use. 

To illustrate this operation, we have selected the project datapro4j\footnote{\emph{Datapro4j}, available from \url{http://www.jrromero.net/tool_datapro4j.html} (accessed June 22, 2020)}, a Java library for processing and handling data from heterogeneous data sources, which contains implementations of the Singleton pattern. Fig.~\ref{fig:exRules} shows two illustrative rules returned by G3P4DPD: one describing positive samples ---consequent \textit{aPattern}--- and one for negative samples ---consequent \textit{notAPattern}. On the one hand, the first rule implies that implementations of the Singleton pattern scrutinised from this repository contain a non-public constructor, and a property of their own type to ensure that only one instance is created. In addition, DIT values are not greater than one, meaning that Singleton instances have no superclasses in this repository, except for \emph{Object}, as any Java class. On the other hand, the second rule determines that those classes with a public constructor whose double invocation is not controlled by exceptions are not a valid implementation of Singleton. Notice that the use of exceptions is an alternative implementation of Singleton, which is usually coded in terms of a private constructor whose invocation is controlled by a static variable. However, the use of design microstructures as grammar operators allows detecting both cases. For this execution, the support and confidence thresholds were set to 0.01 and 0.7. In addition, the list of operators were limited to only those that best represent the Singleton pattern (see Section~\ref{subsec:exp1_operators}).

\begin{figure}[!t]
\begin{lstlisting}[basicstyle=\scriptsize\ttfamily, linewidth=7.4cm, xleftmargin=.015\textwidth, xrightmargin=.015\textwidth]
 (*@{\emph{if}}@*) 
   ctorVisibility(singleton) != public 
   (*@{\emph{and}}@*) aggregation(singleton,singleton) != notLinked
   (*@{\emph{and}}@*) DIT(singleton) < 2
 (*@{\emph{then}}@*) 
   aPattern

 (*@{\emph{if}}@*) 
   ctorVisibility(singleton) = public 
   (*@{\emph{and}}@*) controlledExcept(singleton) = false 
   (*@{\emph{and}}@*) controlledInit(singleton) = false 
 (*@{\emph{then}}@*) 
   notAPattern
\end{lstlisting}
\caption{Two sample rules generated by the G3P4DPD algorithm}
\label{fig:exRules}
\end{figure}

\begin{figure}[!t]
\centerline{\includegraphics[scale=0.83]{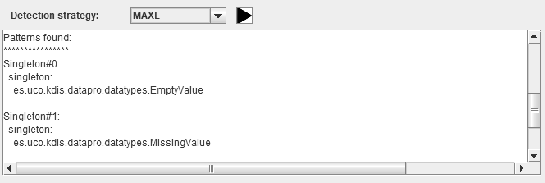}}
\caption{Step 3: Recovery of Singleton pattern instances}
\label{fig:dpdtool-classification}
\end{figure}

Finally, as for the \textit{recovery of design patterns}, the classification strategy has to be selected. For the datapro4j example, we applied the MAXL strategy. Fig.~\ref{fig:dpdtool-classification} shows the implementations found for this pattern and repository, which are highly coincident with the actual specification of the library. It is also worth noting that these implementations could be added to the repository in order to gradually adjust this database to the corporate culture in future detections.

\section{Threats to Validity}\label{sec:threads}

Internal threats are those related to aspects of the experimentation that cannot ensure the causality of the obtained results. Here, the stochastic nature of the algorithm, as well as its setup and parametrisation are internal threats to be considered. Therefore, all experiments are based on 30 independent executions. Furthermore, a parameter study was conducted to determine the best values. As for the construction of the detection model, a stratified 10-fold cross validation is performed to avoid any bias due to the training data. Another threat for the internal validity refers to the setup of the detection method and, more specifically, to the selection of the pruning and classification strategies. On the one hand, the database coverage method has been considered for pruning, as it is a well-known, proved method in the AC literature, and has been already applied jointly to four of the classification strategies considered in this study. On the other hand, the detection performance of several classification strategies has been analysed as part of the experimentation. The statistical analysis has shown that the detection model is not greatly affected by these algorithms.

External validity is related to the generalisation of the experimental results. In this approach, we have selected 15 design patterns for validation and comparison purposes. In addition, these patterns present different number of roles and multiple roles played by a single artefact. Even so, the low number of training samples available for some DPs could limit the generalisation of conclusions. Similarly, the use of DPB and P-Mart repositories, which contain Java projects, implies that our conclusions could not be directly extrapolated to organisational environments of different nature. Nevertheless, these projects are of non-trivial size and they have become frequent benchmarks within the field. The application of \ourtechnique to other industrial projects might require its adaptation to their specific requirements, \eg other programming languages or design microstructures (grammar operators). In this sense, the possibility to extend the collection of operators and the flexibility provided by the CFG makes \ourtechnique adaptable to organisational changes.

\section{Concluding Remarks}\label{sec:concluding}

\ourtechnique was introduced as a novel automatic approach for design pattern detection based on evolutionary machine learning. Knowledge from code repositories is extracted by means of G3P4DPD in form of association rules, a highly readable format to represent knowledge~\citep{Grosan11}. The use of an extendable context-free grammar to declare the syntax of rules makes the learning process highly flexible and adaptable to new organisational environments and design patterns. The application of a pruning method and several classification strategies --- already proved in well-known associative classification approaches --- to select those rules with best detection capabilities leads \ourtechnique to accurate and robust predictions.

An extensive experimentation shows that \ourtechnique is able to generate high-quality rules describing implementations of structural, creational and behavioural design patterns. A first study reveals that the prediction performance remains robust --- also improving in general terms to other related proposal --- even when a single common parametrisation is used for all the design patterns, without the need to adjust it individually. We think that this fact could significantly increase its application in practice, since the software engineer is not required to adjust the parameters. We have also analysed whether the detection capabilities of \ourtechnique are affected by changes in the training conditions, using a total of 15 DPs, including behavioural, creational and structural patterns. This is the largest set of patterns analysed so far for a ML-based DPD proposal. Results reveal that \ourtechnique remains highly competitive even when very few DP examples are available for learning. Compared to reference DPD methods and tools, \ourtechnique supports additional DPs and detects more implementations, although the low number of training samples could cause some variability in the rate of false positives. A demonstration tool is provided to show its use and allow programmers to analyse their own Java projects.

In the future we plan to incorporate new design microstructures --- in form of grammar operators --- that facilitate the detection of the remaining DPs. In addition, the current tool can be evolved and integrated within existing IDE platforms like Eclipse, thus integrating detection capabilities as part of the programming and maintenance tasks.

\section*{Additional Material}
\label{sec:additional-material}

For replicability purposes, experimentation data, such as the generated DP instances, are available for download, as well as the results from the experimentation and statistical analysis, and the demonstration tool, from 
\url{https://www.uco.es/kdis/sbse/geml/}

\section*{Acknowledgements}
This work was supported by the Spanish Ministry of Economy and Competitiveness [project TIN2017-83445-P], the Spanish Ministry of Education under the FPU program [grant FPU17/00799] and the University of C\'ordoba [postdoctoral grant ``Plan propio - mod. 2.4''].

\appendix

\section{Extended results for the parameter study}

Tables~\ref{tab:singleton-final}-~\ref{tab:composite-final} compile the results for all combinations of classification strategy, support and confidence thresholds, for the Singleton, Adapter, Factory Method and Composite patterns, respectively. Values represent the average and the standard deviation of each performance measure, i.e. accuracy, precision, recall, specificity and $F_1$. Bold typeface is used to highlight the best result for the respective classification strategy, and shaded cells represent the global best value for each measure.

\begin{table*}[!t]
\caption{Classification performance for the Singleton}
\label{tab:singleton-final}
\centering
\scriptsize{
\begin{tabular}{c c c c c c c}
\hline
& Supp - Conf & Accuracy & Precision & Recall & Specificity & $F_1$ \\ 
 \hline
 
\multirow{ 9}{*}{\rotatebox{90}{MAXL}}
& (0.01) - (0.5) & $ 0.9395 \pm 0.0140 $ & $ 0.9436 \pm 0.0164 $ & $ 0.9008 \pm 0.0311 $ & $ 0.9631 \pm 0.0115 $ & $ 0.9152 \pm 0.0222 $ \\
& (0.01) - (0.6) & $ 0.9434 \pm 0.0125 $ & $ 0.9524 \pm 0.0163 $ & $ 0.9029 \pm 0.0262 $ & $ 0.9681 \pm 0.0111 $ & $ 0.9200 \pm 0.0187 $ \\
& (0.01) - (0.7) & $ \bm{0.9458 \pm 0.0123} $ & $ \bm{0.9535 \pm 0.0140} $ & $ \bm{0.9084 \pm 0.0288} $ & $ \bm{0.9685 \pm 0.0106} $ & $\bm{0.9246 \pm 0.0184} $ \\
& (0.05) - (0.5) & $ 0.9317 \pm 0.0171 $ & $ 0.9427 \pm 0.0195 $ & $ 0.8821 \pm 0.0327 $ & $ 0.9617 \pm 0.0137 $ & $ 0.9038 \pm 0.0265 $ \\
& (0.05) - (0.6) & $ 0.9190 \pm 0.0142 $ & $ 0.9434 \pm 0.0200 $ & $ 0.8441 \pm 0.0279 $ & $ 0.9641 \pm 0.0135 $ & $ 0.8820 \pm 0.0208 $ \\
& (0.05) - (0.7) & $ 0.9197 \pm 0.0122 $ & $ 0.9422 \pm 0.0195 $ & $ 0.8467 \pm 0.0228 $ & $ 0.9640 \pm 0.0130 $ & $ 0.8832 \pm 0.0187 $ \\
& (0.1) - (0.5) & $ 0.9373 \pm 0.0140 $ & $ 0.9461 \pm 0.0175 $ & $ 0.8916 \pm 0.0293 $ & $ 0.9652 \pm 0.0104 $ & $ 0.9119 \pm 0.0206 $ \\
& (0.1) - (0.6) & $ 0.9254 \pm 0.0107 $ & $ 0.9493 \pm 0.0175 $ & $ 0.8550 \pm 0.0230 $ & $ 0.9678 \pm 0.0105 $ & $ 0.8912 \pm 0.0165 $ \\
& (0.1) - (0.7) & $ 0.9225 \pm 0.0086 $ & $ 0.9458 \pm 0.0148 $ & $ 0.8508 \pm 0.0183 $ & $ 0.9660 \pm 0.0106 $ & $ 0.8879 \pm 0.0117 $ \\
\hline

\multirow{ 9}{*}{\rotatebox{90}{DFML}}
& (0.01) - (0.5) & $ 0.9431 \pm 0.0132 $ & $ 0.9495 \pm 0.0134 $ & $ 0.9038 \pm 0.0353 $ & $ 0.9670 \pm 0.0097 $ & $ 0.9197 \pm 0.0213 $ \\
& (0.01) - (0.6) & $ 0.9454 \pm 0.0118 $ & $ 0.9493 \pm 0.0143 $ & $ 0.9126 \pm 0.0234 $ & $ 0.9654 \pm 0.0105 $ & $ 0.9239 \pm 0.0173 $ \\
& (0.01) - (0.7) & $ \bm{0.9490 \pm 0.0143} $ & $\bm{0.9539 \pm 0.0143}$ & $ 0.9162 \pm 0.0309 $ & $\bm{0.9689 \pm 0.0110}$ & $\bm{0.9290 \pm 0.0220} $ \\
& (0.05) - (0.5) & $ 0.9291 \pm 0.0186 $ & $ 0.9166 \pm 0.0299 $ & $ 0.9083 \pm 0.0288 $ & $ 0.9418 \pm 0.0237 $ & $ 0.9054 \pm 0.0242 $ \\
& (0.05) - (0.6) & $ 0.9229 \pm 0.0129 $ & $ 0.9370 \pm 0.0219 $ & $ 0.8614 \pm 0.0195 $ & $ 0.9599 \pm 0.0147 $ & $ 0.8897 \pm 0.0179 $ \\
& (0.05) - (0.7) & $ 0.9223 \pm 0.0123 $ & $ 0.9382 \pm 0.0226 $ & $ 0.8586 \pm 0.0171 $ & $ 0.9612 \pm 0.0152 $ & $ 0.8894 \pm 0.0165 $ \\
& (0.1) - (0.5) & $ 0.9204 \pm 0.0167 $ & $ 0.8906 \pm 0.0268 $ & $ \bm{0.9189 \pm 0.0214} $ & $ 0.9214 \pm 0.0219 $ & $ 0.8980 \pm 0.0202 $ \\
& (0.1) - (0.6) & $ 0.9263 \pm 0.0116 $ & $ 0.9350 \pm 0.0213 $ & $ 0.8754 \pm 0.0211 $ & $ 0.9569 \pm 0.0154 $ & $ 0.8961 \pm 0.0166 $ \\
& (0.1) - (0.7) & $ 0.9266 \pm 0.0088 $ & $ 0.9414 \pm 0.0167 $ & $ 0.8669 \pm 0.0132 $ & $ 0.9628 \pm 0.0118 $ & $ 0.8952 \pm 0.0116 $ \\
\hline

\multirow{ 9}{*}{\rotatebox{90}{DFML$_{\chi^2}$}}
& (0.01) - (0.5) & $ 0.9429 \pm 0.0116 $ & $ 0.9320 \pm 0.0185 $ & $ 0.9266 \pm 0.0296 $ & $ 0.9531 \pm 0.0144 $ & $ 0.9227 \pm 0.0175 $ \\
& (0.01) - (0.6) & $ 0.9520 \pm 0.0112 $ & $ 0.9409 \pm 0.0145 $ & $ 0.9407 \pm 0.0238 $ & $ 0.9589 \pm 0.0113 $ & $ 0.9351 \pm 0.0173 $ \\
& (0.01) - (0.7) & \cellcolor{black!25}$ \bm{0.9561 \pm 0.0136} $ & $ \bm{0.9460 \pm 0.0147} $ & \cellcolor{black!25}$ \bm{0.9461 \pm 0.0298} $ & $ 0.9621 \pm 0.0118 $ & \cellcolor{black!25}$ \bm{0.9411 \pm 0.0202} $ \\
& (0.05) - (0.5) & $ 0.9404 \pm 0.0117 $ & $ 0.9346 \pm 0.0163 $ & $ 0.9181 \pm 0.0249 $ & $ 0.9539 \pm 0.0132 $ & $ 0.9192 \pm 0.0175 $ \\
& (0.05) - (0.6) & $ 0.9253 \pm 0.0106 $ & $ 0.9347 \pm 0.0183 $ & $ 0.8712 \pm 0.0199 $ & $ 0.9577 \pm 0.0126 $ & $ 0.8943 \pm 0.0157 $ \\
& (0.05) - (0.7) & $ 0.9220 \pm 0.0106 $ & $ 0.9295 \pm 0.0202 $ & $ 0.8680 \pm 0.0184 $ & $ 0.9550 \pm 0.0134 $ & $ 0.8909 \pm 0.0151 $ \\
& (0.1) - (0.5) & $ 0.9433 \pm 0.0116 $ & $ 0.9372 \pm 0.0192 $ & $ 0.9207 \pm 0.0186 $ & $ 0.9571 \pm 0.0130 $ & $ 0.9234 \pm 0.0162 $ \\
& (0.1) - (0.6) & $ 0.9282 \pm 0.0103 $ & $ 0.9359 \pm 0.0187 $ & $ 0.8779 \pm 0.0169 $ & $ 0.9586 \pm 0.0121 $ & $ 0.8988 \pm 0.0145 $ \\
& (0.1) - (0.7) & $ 0.9284 \pm 0.0094 $ & $ 0.9420 \pm 0.0151 $ & $ 0.8726 \pm 0.0156 $ & $ \bm{0.9623 \pm 0.0103} $ & $ 0.8979 \pm 0.0131 $ \\
\hline

\multirow{ 9}{*}{\rotatebox{90}{DFML$_{Lap}$}}
& (0.01) - (0.5) & $ 0.9278 \pm 0.0129 $ & $ 0.9430 \pm 0.0136 $ & $ 0.8690 \pm 0.0317 $ & $ 0.9635 \pm 0.0084 $ & $ 0.8960 \pm 0.0216 $ \\
& (0.01) - (0.6) & $ 0.9370 \pm 0.0130 $ & $ 0.9489 \pm 0.0145 $ & $ 0.8894 \pm 0.0256 $ & $ 0.9660 \pm 0.0107 $ & $ 0.9103 \pm 0.0193 $ \\
& (0.01) - (0.7) & $ \bm{0.9422 \pm 0.0130} $ & $ 0.9498 \pm 0.0174 $ & $ \bm{0.9028 \pm 0.0275} $ & $ 0.9661 \pm 0.0127 $ & $ \bm{0.9191 \pm 0.0192} $ \\
& (0.05) - (0.5) & $ 0.9265 \pm 0.0134 $ & $ 0.9487 \pm 0.0195 $ & $ 0.8613 \pm 0.0282 $ & $ 0.9658 \pm 0.0136 $ & $ 0.8933 \pm 0.0212 $ \\
& (0.05) - (0.6) & $ 0.9125 \pm 0.0133 $ & $ 0.9448 \pm 0.0201 $ & $ 0.8240 \pm 0.0270 $ & $ 0.9657 \pm 0.0130 $ & $ 0.8707 \pm 0.0210 $ \\
& (0.05) - (0.7) & $ 0.9189 \pm 0.0126 $ & $ 0.9452 \pm 0.0196 $ & $ 0.8400 \pm 0.0244 $ & $ 0.9668 \pm 0.0114 $ & $ 0.8815 \pm 0.0195 $ \\
& (0.1) - (0.5) & $ 0.9329 \pm 0.0129 $ & $ 0.9529 \pm 0.0165 $ & $ 0.8719 \pm 0.0308 $ & $ 0.9700 \pm 0.0094 $ & $ 0.9032 \pm 0.0199 $ \\
& (0.1) - (0.6) & $ 0.9198 \pm 0.0111 $ & \cellcolor{black!25}$ \bm{0.9542 \pm 0.0136} $ & $ 0.8343 \pm 0.0263 $ & $ 0.9711 \pm 0.0084 $ & $ 0.8803 \pm 0.0180 $ \\
& (0.1) - (0.7) & $ 0.9182 \pm 0.0111 $ & $ 0.9532 \pm 0.0127 $ & $ 0.8307 \pm 0.0275 $ & \cellcolor{black!25}$ \bm{0.9713 \pm 0.0084} $ & $ 0.8785 \pm 0.0175 $ \\
\hline

\hline
\end{tabular}
}
\end{table*}

\begin{table*}[!t]
\caption{Classification performance for the Adapter}
\label{tab:adapter-final}
\centering
\scriptsize{
\begin{tabular}{c c c c c c c}
\hline
& Supp - Conf & Accuracy & Precision & Recall & Specificity & $F_1$ \\ 
 \hline
 
\multirow{ 9}{*}{\rotatebox{90}{MAXL}}
& (0.01) - (0.5) & $ 0.8542 \pm 0.0047 $ & $ 0.8356 \pm 0.0051 $ & $ 0.8885 \pm 0.0066 $ & $ 0.8190 \pm 0.0059 $ & $ 0.8604 \pm 0.0047 $ \\
& (0.01) - (0.6) & $ 0.8624 \pm 0.0034 $ & $ 0.8402 \pm 0.0035 $ & $ 0.9021 \pm 0.0058 $ & $ 0.8218 \pm 0.0044 $ & $ 0.8692 \pm 0.0034 $ \\
& (0.01) - (0.7) & $ 0.8657 \pm 0.0028 $ & $ \bm{0.8431 \pm 0.0029} $ & $ 0.9048 \pm 0.0047 $ & $ \bm{0.8258 \pm 0.0037} $ & $ 0.8722 \pm 0.0028 $ \\
& (0.05) - (0.5) & $ 0.8604 \pm 0.0024 $ & $ 0.8411 \pm 0.0023 $ & $ 0.8953 \pm 0.0039 $ & $ 0.8247 \pm 0.0027 $ & $ 0.8666 \pm 0.0024 $ \\
& (0.05) - (0.6) & $ 0.8630 \pm 0.0025 $ & $ 0.8411 \pm 0.0029 $ & $ 0.9012 \pm 0.0046 $ & $ 0.8237 \pm 0.0039 $ & $ 0.8695 \pm 0.0025 $ \\
& (0.05) - (0.7) & $ 0.8642 \pm 0.0026 $ & $ 0.8419 \pm 0.0029 $ & $ 0.9028 \pm 0.0047 $ & $ 0.8246 \pm 0.0040 $ & $ 0.8706 \pm 0.0026 $ \\
& (0.1) - (0.5) & $ 0.8667 \pm 0.0015 $ & $ 0.8402 \pm 0.0013 $ & $ 0.9121 \pm 0.0029 $ & $ 0.8203 \pm 0.0013 $ & $ 0.8740 \pm 0.0017 $ \\
& (0.1) - (0.6) & $ 0.8672 \pm 0.0018 $ & $ 0.8407 \pm 0.0021 $ & $ \bm{0.9124 \pm 0.0024} $ & $ 0.8208 \pm 0.0026 $ & $ 0.8744 \pm 0.0017 $ \\
& (0.1) - (0.7) & $ \bm{0.8675 \pm 0.0019} $ & $ 0.8413 \pm 0.0030 $ & $ 0.9120 \pm 0.0024 $ & $ 0.8220 \pm 0.0032 $ & $ \bm{0.8746 \pm 0.0018} $ \\
\hline

\multirow{ 9}{*}{\rotatebox{90}{DFML}}
& (0.01) - (0.5) & $ 0.8621 \pm 0.0045 $ & $ 0.8392 \pm 0.0046 $ & $ 0.9020 \pm 0.0073 $ & $ 0.8211 \pm 0.0061 $ & $ 0.8686 \pm 0.0046 $ \\
& (0.01) - (0.6) & $ 0.8641 \pm 0.0041 $ & \cellcolor{black!25}$ \bm{0.8464 \pm 0.0044} $ & $ 0.8964 \pm 0.0083 $ &\cellcolor{black!25} $ \bm{0.8309 \pm 0.0059} $ & $ 0.8697 \pm 0.0044 $ \\
& (0.01) - (0.7) & $ \bm{0.8679 \pm 0.0028} $ & $ 0.8450 \pm 0.0040 $ & $ 0.9069 \pm 0.0057 $ & $ 0.8279 \pm 0.0059 $ & $ \bm{0.8743 \pm 0.0028} $ \\
& (0.05) - (0.5) & $ 0.8636 \pm 0.0028 $ & $ 0.8418 \pm 0.0022 $ & $ 0.9021 \pm 0.0071 $ & $ 0.8242 \pm 0.0036 $ & $ 0.8701 \pm 0.0033 $ \\
& (0.05) - (0.6) & $ 0.8634 \pm 0.0026 $ & $ 0.8447 \pm 0.0044 $ & $ 0.8969 \pm 0.0069 $ & $ 0.8292 \pm 0.0061 $ & $ 0.8692 \pm 0.0029 $ \\
& (0.05) - (0.7) & $ 0.8665 \pm 0.0033 $ & $ 0.8428 \pm 0.0037 $ & $ 0.9069 \pm 0.0060 $ & $ 0.8250 \pm 0.0051 $ & $ 0.8730 \pm 0.0033 $ \\
& (0.1) - (0.5) & $ 0.8655 \pm 0.0024 $ & $ 0.8406 \pm 0.0024 $ & $ 0.9084 \pm 0.0054 $ & $ 0.8214 \pm 0.0031 $ & $ 0.8724 \pm 0.0026 $ \\
& (0.1) - (0.6) & $ 0.8632 \pm 0.0033 $ & $ 0.8440 \pm 0.0038 $ & $ 0.8974 \pm 0.0073 $ & $ 0.8283 \pm 0.0057 $ & $ 0.8691 \pm 0.0035 $ \\
& (0.1) - (0.7) & $ 0.8662 \pm 0.0026 $ & $ 0.8414 \pm 0.0032 $ & $ \bm{0.9084 \pm 0.0047} $ & $ 0.8230 \pm 0.0036 $ & $ 0.8730 \pm 0.0028 $ \\
\hline

\multirow{ 9}{*}{\rotatebox{90}{DFML$_{\chi^2}$}}
& (0.01) - (0.5) & $ 0.8673 \pm 0.0013 $ & $ 0.8401 \pm 0.0018 $ & $ 0.9134 \pm 0.0017 $ & $ 0.8201 \pm 0.0019 $ & $ 0.8746 \pm 0.0013 $ \\
& (0.01) - (0.6) & $ 0.8680 \pm 0.0022 $ & $ 0.8423 \pm 0.0027 $ & $ 0.9119 \pm 0.0033 $ & $ 0.8229 \pm 0.0030 $ & $ 0.8750 \pm 0.0022 $ \\
& (0.01) - (0.7) & \cellcolor{black!25}$ \bm{0.8688 \pm 0.0022} $ & $ \bm{0.8430 \pm 0.0028} $ & $ 0.9121 \pm 0.0032 $ & $ \bm{0.8244 \pm 0.0038} $ & \cellcolor{black!25}$ \bm{0.8757 \pm 0.0020} $ \\
& (0.05) - (0.5) & $ 0.8670 \pm 0.0013 $ & $ 0.8400 \pm 0.0015 $ & $ 0.9130 \pm 0.0018 $ & $ 0.8198 \pm 0.0016 $ & $ 0.8744 \pm 0.0013 $ \\
& (0.05) - (0.6) & $ 0.8671 \pm 0.0019 $ & $ 0.8406 \pm 0.0025 $ & $ 0.9119 \pm 0.0024 $ & $ 0.8212 \pm 0.0032 $ & $ 0.8742 \pm 0.0018 $ \\
& (0.05) - (0.7) & $ 0.8684 \pm 0.0016 $ & $ 0.8421 \pm 0.0027 $ & $ 0.9129 \pm 0.0026 $ & $ 0.8229 \pm 0.0036 $ & $ 0.8754 \pm 0.0014 $ \\
& (0.1) - (0.5) & $ 0.8674 \pm 0.0004 $ & $ 0.8401 \pm 0.0013 $ & \cellcolor{black!25}$ \bm{0.9137 \pm 0.0010} $ & $ 0.8198 \pm 0.0010 $ & $ 0.8748 \pm 0.0009 $ \\
& (0.1) - (0.6) & $ 0.8676 \pm 0.0016 $ & $ 0.8407 \pm 0.0021 $ & $ 0.9133 \pm 0.0017 $ & $ 0.8208 \pm 0.0025 $ & $ 0.8748 \pm 0.0014 $ \\
& (0.1) - (0.7) & $ 0.8679 \pm 0.0019 $ & $ 0.8414 \pm 0.0030 $ & $ 0.9127 \pm 0.0022 $ & $ 0.8220 \pm 0.0032 $ & $ 0.8750 \pm 0.0018 $ \\
\hline

\multirow{ 9}{*}{\rotatebox{90}{DFML$_{Lap}$}}
& (0.01) - (0.5) & $ 0.8544 \pm 0.0038 $ & $ 0.8388 \pm 0.0035 $ & $ 0.8840 \pm 0.0070 $ & $ 0.8241 \pm 0.0044 $ & $ 0.8600 \pm 0.0040 $ \\
& (0.01) - (0.6) & $ 0.8629 \pm 0.0031 $ & $ 0.8416 \pm 0.0032 $ & $ 0.9008 \pm 0.0063 $ & $ 0.8240 \pm 0.0043 $ & $ 0.8694 \pm 0.0034 $ \\
& (0.01) - (0.7) & $ 0.8656 \pm 0.0028 $ & $ \bm{0.8442 \pm 0.0027} $ & $ 0.9025 \pm 0.0046 $ & $ \bm{0.8277 \pm 0.0034} $ & $ 0.8718 \pm 0.0028 $ \\
& (0.05) - (0.5) & $ 0.8603 \pm 0.0025 $ & $ 0.8410 \pm 0.0023 $ & $ 0.8950 \pm 0.0040 $ & $ 0.8246 \pm 0.0027 $ & $ 0.8664 \pm 0.0026 $ \\
& (0.05) - (0.6) & $ 0.8638 \pm 0.0024 $ & $ 0.8416 \pm 0.0029 $ & $ 0.9023 \pm 0.0050 $ & $ 0.8243 \pm 0.0039 $ & $ 0.8703 \pm 0.0025 $ \\
& (0.05) - (0.7) & $ 0.8641 \pm 0.0024 $ & $ 0.8423 \pm 0.0027 $ & $ 0.9020 \pm 0.0050 $ & $ 0.8252 \pm 0.0038 $ & $ 0.8704 \pm 0.0025 $ \\
& (0.1) - (0.5) & $ 0.8666 \pm 0.0015 $ & $ 0.8402 \pm 0.0013 $ & $ 0.9119 \pm 0.0028 $ & $ 0.8202 \pm 0.0012 $ & $ 0.8739 \pm 0.0017 $ \\
& (0.1) - (0.6) & $ 0.8672 \pm 0.0019 $ & $ 0.8407 \pm 0.0022 $ & $ \bm{0.9124 \pm 0.0024} $ & $ 0.8209 \pm 0.0027 $ & $ \bm{0.8744 \pm 0.0018} $ \\
& (0.1) - (0.7) & $ \bm{0.8673 \pm 0.0019} $ & $ 0.8413 \pm 0.0029 $ & $ 0.9113 \pm 0.0027 $ & $ 0.8221 \pm 0.0031 $ & $ 0.8743 \pm 0.0019 $ \\
\hline

\hline
\end{tabular}
}
\end{table*}

\begin{table*}[!t]
\caption{Classification performance for the Factory Method}
\label{tab:fMethod-final}
\centering
\scriptsize{
\begin{tabular}{c c c c c c c}
\hline
& Supp - Conf & Accuracy & Precision & Recall & Specificity & $F_1$ \\ 
 \hline
 
\multirow{ 9}{*}{\rotatebox{90}{MAXL}}
& (0.01) - (0.5) & $ 0.8200 \pm 0.0086 $ & $ \bm{0.8331 \pm 0.0089} $ & $ 0.8348 \pm 0.0099 $ & $ 0.8028 \pm 0.0123 $ & $ 0.8329 \pm 0.0079 $ \\
& (0.01) - (0.6) & $ 0.8198 \pm 0.0081 $ & $ 0.8294 \pm 0.0088 $ & $ 0.8404 \pm 0.0110 $ & $ 0.7957 \pm 0.0127 $ & $ 0.8335 \pm 0.0077 $ \\
& (0.01) - (0.7) & $ \bm{0.8266 \pm 0.0081} $ & $ 0.8311 \pm 0.0082 $ & $ 0.8540 \pm 0.0135 $ & $ 0.7946 \pm 0.0119 $ & $ 0.8411 \pm 0.0081 $ \\
& (0.05) - (0.5) & $ 0.7946 \pm 0.0074 $ & $ 0.8313 \pm 0.0088 $ & $ 0.7790 \pm 0.0132 $ & $ \bm{0.8128 \pm 0.0136} $ & $ 0.8026 \pm 0.0076 $ \\
& (0.05) - (0.6) & $ 0.8028 \pm 0.0098 $ & $ 0.8257 \pm 0.0102 $ & $ 0.8062 \pm 0.0156 $ & $ 0.7988 \pm 0.0142 $ & $ 0.8141 \pm 0.0099 $ \\
& (0.05) - (0.7) & $ 0.8156 \pm 0.0079 $ & $ 0.8260 \pm 0.0076 $ & $ 0.8356 \pm 0.0173 $ & $ 0.7922 \pm 0.0131 $ & $ 0.8292 \pm 0.0085 $ \\
& (0.1) - (0.5) & $ 0.8102 \pm 0.0110 $ & $ 0.8070 \pm 0.0128 $ & $ 0.8556 \pm 0.0117 $ & $ 0.7573 \pm 0.0216 $ & $ 0.8291 \pm 0.0091 $ \\
& (0.1) - (0.6) & $ 0.8174 \pm 0.0087 $ & $ 0.8095 \pm 0.0087 $ & $ 0.8691 \pm 0.0126 $ & $ 0.7572 \pm 0.0148 $ & $ 0.8366 \pm 0.0081 $ \\
& (0.1) - (0.7) & $ 0.8221 \pm 0.0116 $ & $ 0.7998 \pm 0.0122 $ & $ \bm{0.8992 \pm 0.0130} $ & $ 0.7322 \pm 0.0213 $ & $ \bm{0.8451 \pm 0.0097} $ \\
\hline

\multirow{ 9}{*}{\rotatebox{90}{DFML}}
& (0.01) - (0.5) & $ 0.8124 \pm 0.0089 $ & $ 0.8013 \pm 0.0096 $ & $ 0.8697 \pm 0.0121 $ & $ 0.7456 \pm 0.0159 $ & $ 0.8328 \pm 0.0080 $ \\
& (0.01) - (0.6) & $ 0.8198 \pm 0.0099 $ & $ 0.8141 \pm 0.0118 $ & $ 0.8663 \pm 0.0157 $ & $ 0.7656 \pm 0.0186 $ & $ 0.8378 \pm 0.0092 $ \\
& (0.01) - (0.7) & $ \bm{0.8275 \pm 0.0077} $ & $ \bm{0.8177 \pm 0.0075} $ & $ 0.8783 \pm 0.0175 $ & $ \bm{0.7683 \pm 0.0132} $ & $ \bm{0.8455 \pm 0.0082} $ \\
& (0.05) - (0.5) & $ 0.7958 \pm 0.0091 $ & $ 0.7926 \pm 0.0115 $ & $ 0.8457 \pm 0.0126 $ & $ 0.7376 \pm 0.0189 $ & $ 0.8166 \pm 0.0079 $ \\
& (0.05) - (0.6) & $ 0.8069 \pm 0.0109 $ & $ 0.8047 \pm 0.0099 $ & $ 0.8505 \pm 0.0163 $ & $ 0.7560 \pm 0.0150 $ & $ 0.8253 \pm 0.0105 $ \\
& (0.05) - (0.7) & $ 0.8219 \pm 0.0080 $ & $ 0.8115 \pm 0.0084 $ & $ 0.8753 \pm 0.0149 $ & $ 0.7597 \pm 0.0142 $ & $ 0.8408 \pm 0.0080 $ \\
& (0.1) - (0.5) & $ 0.7908 \pm 0.0089 $ & $ 0.7666 \pm 0.0098 $ & $ 0.8853 \pm 0.0144 $ & $ 0.6805 \pm 0.0185 $ & $ 0.8197 \pm 0.0081 $ \\
& (0.1) - (0.6) & $ 0.8062 \pm 0.0099 $ & $ 0.7841 \pm 0.0102 $ & $ 0.8886 \pm 0.0137 $ & $ 0.7100 \pm 0.0171 $ & $ 0.8315 \pm 0.0088 $ \\
& (0.1) - (0.7) & $ 0.8143 \pm 0.0145 $ & $ 0.7887 \pm 0.0125 $ & $ \bm{0.9001 \pm 0.0175} $ & $ 0.7143 \pm 0.0194 $ & $ 0.8393 \pm 0.0130 $ \\
\hline

\multirow{ 9}{*}{\rotatebox{90}{DFML$_{\chi^2}$}}
& (0.01) - (0.5) & $ 0.8201 \pm 0.0082 $ & $ \bm{0.8202 \pm 0.0092} $ & $ 0.8561 \pm 0.0113 $ & $ \bm{0.7780 \pm 0.0151} $ & $ 0.8363 \pm 0.0074 $ \\
& (0.01) - (0.6) & $ 0.8221 \pm 0.0090 $ & $ 0.8180 \pm 0.0094 $ & $ 0.8655 \pm 0.0176 $ & $ 0.7716 \pm 0.0153 $ & $ 0.8392 \pm 0.0090 $ \\
& (0.01) - (0.7) & \cellcolor{black!25}$ \bm{0.8308 \pm 0.0089} $ & $ 0.8180 \pm 0.0071 $ & $ 0.8863 \pm 0.0218 $ & $ 0.7661 \pm 0.0135 $ & $ 0.8489 \pm 0.0099 $ \\
& (0.05) - (0.5) & $ 0.8102 \pm 0.0096 $ & $ 0.8120 \pm 0.0084 $ & $ 0.8466 \pm 0.0160 $ & $ 0.7677 \pm 0.0133 $ & $ 0.8271 \pm 0.0095 $ \\
& (0.05) - (0.6) & $ 0.8197 \pm 0.0117 $ & $ 0.8105 \pm 0.0098 $ & $ 0.8725 \pm 0.0195 $ & $ 0.7582 \pm 0.0153 $ & $ 0.8382 \pm 0.0114 $ \\
& (0.05) - (0.7) & $ 0.8304 \pm 0.0082 $ & $ 0.8113 \pm 0.0081 $ & $ 0.8965 \pm 0.0158 $ & $ 0.7533 \pm 0.0145 $ & \cellcolor{black!25}$ \bm{0.8503 \pm 0.0081} $ \\
& (0.1) - (0.5) & $ 0.8134 \pm 0.0102 $ & $ 0.7910 \pm 0.0111 $ & $ 0.8943 \pm 0.0119 $ & $ 0.7190 \pm 0.0205 $ & $ 0.8377 \pm 0.0083 $ \\
& (0.1) - (0.6) & $ 0.8227 \pm 0.0088 $ & $ 0.7971 \pm 0.0093 $ & $ 0.9046 \pm 0.0124 $ & $ 0.7271 \pm 0.0163 $ & $ 0.8461 \pm 0.0076 $ \\
& (0.1) - (0.7) & $ 0.8243 \pm 0.0111 $ & $ 0.7940 \pm 0.0123 $ & \cellcolor{black!25}$ \bm{0.9154 \pm 0.0134} $ & $ 0.7180 \pm 0.0228 $ & $ 0.8490 \pm 0.0091 $ \\
\hline

\multirow{ 9}{*}{\rotatebox{90}{DFML$_{Lap}$}}
& (0.01) - (0.5) & $ 0.8097 \pm 0.0076 $ & \cellcolor{black!25}$ \bm{0.8442 \pm 0.0066} $ & $ 0.7957 \pm 0.0125 $ & $ 0.8262 \pm 0.0090 $ & $ 0.8177 \pm 0.0079 $ \\
& (0.01) - (0.6) & $ 0.8097 \pm 0.0101 $ & $ 0.8325 \pm 0.0102 $ & $ 0.8120 \pm 0.0137 $ & $ 0.8070 \pm 0.0141 $ & $ 0.8206 \pm 0.0098 $ \\
& (0.01) - (0.7) & $ 0.8189 \pm 0.0097 $ & $ 0.8355 \pm 0.0100 $ & $ 0.8298 \pm 0.0149 $ & $ 0.8062 \pm 0.0137 $ & $ 0.8310 \pm 0.0096 $ \\
& (0.05) - (0.5) & $ 0.7857 \pm 0.0094 $ & $ 0.8315 \pm 0.0088 $ & $ 0.7582 \pm 0.0195 $ & \cellcolor{black!25}$ \bm{0.8177 \pm 0.0140} $ & $ 0.7911 \pm 0.0110 $ \\
& (0.05) - (0.6) & $ 0.7984 \pm 0.0093 $ & $ 0.8280 \pm 0.0114 $ & $ 0.7934 \pm 0.0156 $ & $ 0.8042 \pm 0.0160 $ & $ 0.8083 \pm 0.0096 $ \\
& (0.05) - (0.7) & $ 0.8125 \pm 0.0084 $ & $ 0.8281 \pm 0.0099 $ & $ 0.8256 \pm 0.0178 $ & $ 0.7973 \pm 0.0159 $ & $ 0.8251 \pm 0.0089 $ \\
& (0.1) - (0.5) & $ 0.8048 \pm 0.0118 $ & $ 0.8018 \pm 0.0149 $ & $ 0.8532 \pm 0.0162 $ & $ 0.7483 \pm 0.0271 $ & $ 0.8245 \pm 0.0097 $ \\
& (0.1) - (0.6) & $ 0.8145 \pm 0.0093 $ & $ 0.8080 \pm 0.0105 $ & $ 0.8657 \pm 0.0156 $ & $ 0.7548 \pm 0.0196 $ & $ 0.8338 \pm 0.0086 $ \\
& (0.1) - (0.7) & $ \bm{0.8208 \pm 0.0113} $ & $ 0.7984 \pm 0.0128 $ & $ \bm{0.8993 \pm 0.0141} $ & $ 0.7292 \pm 0.0238 $ & $ \bm{0.8442 \pm 0.0092} $ \\
\hline

\hline
\end{tabular}
}
\end{table*}

\begin{table*}[!t]
\caption{Classification performance for the Decorator}
\label{tab:decorator-final}
\centering
\scriptsize{
\begin{tabular}{c c c c c c c}
\hline
& Supp - Conf & Accuracy & Precision & Recall & Specificity & $F_1$ \\ 
 \hline
 
\multirow{ 9}{*}{\rotatebox{90}{MAXL}}
& (0.01) - (0.5) & $ 0.7848 \pm 0.0135 $ & $ 0.8466 \pm 0.0288 $ & $ 0.5449 \pm 0.0265 $ & $ 0.9303 \pm 0.0132 $ & $ 0.6448 \pm 0.0236 $ \\
& (0.01) - (0.6) & $ 0.7904 \pm 0.0161 $ & $ 0.8369 \pm 0.0393 $ & $ 0.5605 \pm 0.0369 $ & $ 0.9301 \pm 0.0141 $ & $ 0.6560 \pm 0.0331 $ \\
& (0.01) - (0.7) & $ 0.7973 \pm 0.0168 $ & $ 0.8311 \pm 0.0278 $ & $ \bm{0.5898 \pm 0.0393} $ & $ 0.9232 \pm 0.0129 $ & $ 0.6755 \pm 0.0341 $ \\
& (0.05) - (0.5) & $ 0.7882 \pm 0.0125 $ & $ 0.8753 \pm 0.0335 $ & $ 0.5235 \pm 0.0301 $ & $ 0.9489 \pm 0.0134 $ & $ 0.6357 \pm 0.0279 $ \\
& (0.05) - (0.6) & $ 0.7936 \pm 0.0130 $ & $ \bm{0.8778 \pm 0.0271} $ & $ 0.5388 \pm 0.0312 $ & $ 0.9484 \pm 0.0125 $ & $ 0.6530 \pm 0.0252 $ \\
& (0.05) - (0.7) & $ \bm{0.8051 \pm 0.0141} $ & $ 0.8717 \pm 0.0236 $ & $ 0.5779 \pm 0.0364 $ & $ 0.9431 \pm 0.0105 $ & $ \bm{0.6794 \pm 0.0287} $ \\
& (0.1) - (0.5) & $ 0.7690 \pm 0.0155 $ & $ 0.8679 \pm 0.0341 $ & $ 0.4693 \pm 0.0379 $ & $ \bm{0.9509 \pm 0.0140} $ & $ 0.5913 \pm 0.0370 $ \\
& (0.1) - (0.6) & $ 0.7725 \pm 0.0147 $ & $ 0.8592 \pm 0.0322 $ & $ 0.4904 \pm 0.0386 $ & $ 0.9438 \pm 0.0148 $ & $ 0.6064 \pm 0.0327 $ \\
& (0.1) - (0.7) & $ 0.7927 \pm 0.0129 $ & $ 0.8526 \pm 0.0307 $ & $ 0.5547 \pm 0.0319 $ & $ 0.9372 \pm 0.0117 $ & $ 0.6584 \pm 0.0275 $ \\
\hline

\multirow{ 9}{*}{\rotatebox{90}{DFML}}
& (0.01) - (0.5) & $ 0.8084 \pm 0.0149 $ & $ 0.8552 \pm 0.0249 $ & $ 0.6135 \pm 0.0327 $ & $ 0.9267 \pm 0.0122 $ & $ 0.6968 \pm 0.0299 $ \\
& (0.01) - (0.6) & $ 0.8128 \pm 0.0149 $ & $ 0.8508 \pm 0.0244 $ & $ 0.6279 \pm 0.0322 $ & $ 0.9253 \pm 0.0141 $ & $ 0.7068 \pm 0.0279 $ \\
& (0.01) - (0.7) & $ 0.8168 \pm 0.0162 $ & $ 0.8393 \pm 0.0239 $ & $ 0.6523 \pm 0.0399 $ & $ 0.9166 \pm 0.0145 $ & $ 0.7203 \pm 0.0328 $ \\
& (0.05) - (0.5) & $ 0.8097 \pm 0.0146 $ & $ 0.8424 \pm 0.0253 $ & $ 0.6242 \pm 0.0375 $ & $ 0.9223 \pm 0.0138 $ & $ 0.7018 \pm 0.0291 $ \\
& (0.05) - (0.6) & $ 0.8104 \pm 0.0147 $ & $ 0.8514 \pm 0.0286 $ & $ 0.6204 \pm 0.0318 $ & $ 0.9260 \pm 0.0136 $ & $ 0.7039 \pm 0.0250 $ \\
& (0.05) - (0.7) & $ \bm{0.8212 \pm 0.0210} $ & $ 0.8438 \pm 0.0310 $ & $ \bm{0.6596 \pm 0.0489} $ & $ 0.9193 \pm 0.0163 $ & $ \bm{0.7262 \pm 0.0386} $ \\
& (0.1) - (0.5) & $ 0.7839 \pm 0.0111 $ & $ \bm{0.8620 \pm 0.0290} $ & $ 0.5219 \pm 0.0273 $ & $ \bm{0.9429 \pm 0.0138} $ & $ 0.6310 \pm 0.0248 $ \\
& (0.1) - (0.6) & $ 0.7851 \pm 0.0166 $ & $ 0.8574 \pm 0.0365 $ & $ 0.5360 \pm 0.0425 $ & $ 0.9363 \pm 0.0162 $ & $ 0.6381 \pm 0.0372 $ \\
& (0.1) - (0.7) & $ 0.8117 \pm 0.0094 $ & $ 0.8479 \pm 0.0239 $ & $ 0.6246 \pm 0.0306 $ & $ 0.9254 \pm 0.0128 $ & $ 0.7061 \pm 0.0210 $ \\
\hline

\multirow{ 9}{*}{\rotatebox{90}{DFML$_{\chi^2}$}}
& (0.01) - (0.5) & $ 0.8143 \pm 0.0174 $ & $ 0.7993 \pm 0.0285 $ & $ 0.7037 \pm 0.0350 $ & $ 0.8815 \pm 0.0189 $ & $ 0.7352 \pm 0.0280 $ \\
& (0.01) - (0.6) & $ 0.8191 \pm 0.0141 $ & $ 0.8042 \pm 0.0258 $ & $ 0.7099 \pm 0.0264 $ & $ 0.8855 \pm 0.0170 $ & $ 0.7416 \pm 0.0229 $ \\
& (0.01) - (0.7) & $ 0.8188 \pm 0.0150 $ & $ 0.7979 \pm 0.0229 $ & $ 0.7184 \pm 0.0340 $ & $ 0.8797 \pm 0.0155 $ & $ 0.7448 \pm 0.0265 $ \\
& (0.05) - (0.5) & $ 0.8118 \pm 0.0141 $ & $ 0.7835 \pm 0.0203 $ & $ 0.7122 \pm 0.0318 $ & $ 0.8722 \pm 0.0167 $ & $ 0.7360 \pm 0.0221 $ \\
& (0.05) - (0.6) & $ 0.8151 \pm 0.0150 $ & $ 0.7993 \pm 0.0248 $ & $ 0.7059 \pm 0.0311 $ & $ 0.8815 \pm 0.0173 $ & $ 0.7384 \pm 0.0238 $ \\
& (0.05) - (0.7) & \cellcolor{black!25}$ \bm{0.8229 \pm 0.0179} $ & $ 0.8043 \pm 0.0269 $ & \cellcolor{black!25}$ \bm{0.7251 \pm 0.0393} $ & $ 0.8824 \pm 0.0171 $ & \cellcolor{black!25}$ \bm{0.7501 \pm 0.0302} $ \\
& (0.1) - (0.5) & $ 0.8009 \pm 0.0123 $ & $ 0.8088 \pm 0.0250 $ & $ 0.6379 \pm 0.0264 $ & $ 0.8997 \pm 0.0157 $ & $ 0.6991 \pm 0.0217 $ \\
& (0.1) - (0.6) & $ 0.8056 \pm 0.0162 $ & $ \bm{0.8208 \pm 0.0266} $ & $ 0.6391 \pm 0.0393 $ & $ \bm{0.9067 \pm 0.0157} $ & $ 0.7031 \pm 0.0304 $ \\
& (0.1) - (0.7) & $ 0.8220 \pm 0.0164 $ & $ 0.8166 \pm 0.0175 $ & $ 0.7014 \pm 0.0377 $ & $ 0.8952 \pm 0.0116 $ & $ 0.7428 \pm 0.0278 $ \\
\hline

\multirow{ 9}{*}{\rotatebox{90}{DFML$_{Lap}$}}
& (0.01) - (0.5) & $ 0.7816 \pm 0.0127 $ & $ 0.8782 \pm 0.0270 $ & $ 0.5031 \pm 0.0246 $ & $ 0.9508 \pm 0.0119 $ & $ 0.6217 \pm 0.0213 $ \\
& (0.01) - (0.6) & $ 0.7833 \pm 0.0150 $ & $ 0.8680 \pm 0.0448 $ & $ 0.5049 \pm 0.0323 $ & $ 0.9524 \pm 0.0128 $ & $ 0.6212 \pm 0.0329 $ \\
& (0.01) - (0.7) & $ 0.7950 \pm 0.0172 $ & $ 0.8699 \pm 0.0306 $ & $ 0.5489 \pm 0.0414 $ & $ 0.9443 \pm 0.0138 $ & $ 0.6549 \pm 0.0382 $ \\
& (0.05) - (0.5) & $ 0.7817 \pm 0.0147 $ & \cellcolor{black!25}$ \bm{0.8850 \pm 0.0410} $ & $ 0.4929 \pm 0.0320 $ & \cellcolor{black!25}$ \bm{0.9570 \pm 0.0150} $ & $ 0.6139 \pm 0.0325 $ \\
& (0.05) - (0.6) & $ 0.7843 \pm 0.0167 $ & $ 0.8838 \pm 0.0393 $ & $ 0.5037 \pm 0.0370 $ & $ 0.9548 \pm 0.0157 $ & $ 0.6251 \pm 0.0322 $ \\
& (0.05) - (0.7) & $ \bm{0.7999 \pm 0.0142} $ & $ 0.8800 \pm 0.0277 $ & $ \bm{0.5547 \pm 0.0336} $ & $ 0.9488 \pm 0.0120 $ & $ \bm{0.6637 \pm 0.0286} $ \\
& (0.1) - (0.5) & $ 0.7614 \pm 0.0137 $ & $ 0.8739 \pm 0.0343 $ & $ 0.4412 \pm 0.0357 $ & $ 0.9559 \pm 0.0131 $ & $ 0.5675 \pm 0.0366 $ \\
& (0.1) - (0.6) & $ 0.7679 \pm 0.0162 $ & $ 0.8722 \pm 0.0372 $ & $ 0.4639 \pm 0.0377 $ & $ 0.9527 \pm 0.0138 $ & $ 0.5870 \pm 0.0361 $ \\
& (0.1) - (0.7) & $ 0.7824 \pm 0.0155 $ & $ 0.8552 \pm 0.0362 $ & $ 0.5174 \pm 0.0356 $ & $ 0.9433 \pm 0.0141 $ & $ 0.6307 \pm 0.0337 $ \\
\hline

\hline
\end{tabular}
}
\end{table*}

\begin{table*}[!t]
\caption{Classification performance for the Composite}
\label{tab:composite-final}
\centering
\scriptsize{
\begin{tabular}{c c c c c c c}
\hline
& Supp - Conf & Accuracy & Precision & Recall & Specificity & $F_1$ \\ 
 \hline
 
\multirow{ 9}{*}{\rotatebox{90}{MAXL}}
& (0.01) - (0.5) & $ 0.8718 \pm 0.0193 $ & $ 0.8201 \pm 0.0791 $ & $ \bm{0.5544 \pm 0.0712} $ & $ 0.9690 \pm 0.0139 $ & $ 0.6309 \pm 0.0668 $ \\
& (0.01) - (0.6) & $ 0.8694 \pm 0.0155 $ & $ 0.8109 \pm 0.1024 $ & $ 0.5533 \pm 0.0491 $ & $ 0.9662 \pm 0.0116 $ & $ 0.6244 \pm 0.0590 $ \\
& (0.01) - (0.7) & $ 0.8626 \pm 0.0152 $ & $ 0.7976 \pm 0.0931 $ & $ 0.5122 \pm 0.0513 $ & $ 0.9700 \pm 0.0151 $ & $ 0.5935 \pm 0.0510 $ \\
& (0.05) - (0.5) & $ 0.8677 \pm 0.0171 $ & $ 0.8211 \pm 0.0795 $ & $ 0.5311 \pm 0.0632 $ & $ 0.9706 \pm 0.0113 $ & $ 0.6145 \pm 0.0577 $ \\
& (0.05) - (0.6) & $ 0.8647 \pm 0.0171 $ & $ 0.8283 \pm 0.0733 $ & $ 0.5178 \pm 0.0569 $ & $ 0.9711 \pm 0.0116 $ & $ 0.6062 \pm 0.0585 $ \\
& (0.05) - (0.7) & $ \bm{0.8737 \pm 0.0179} $ & $ \bm{0.8309 \pm 0.0953} $ & $ 0.5456 \pm 0.0561 $ & $ 0.9742 \pm 0.0112 $ & $ \bm{0.6322 \pm 0.0586} $ \\
& (0.1) - (0.5) & $ 0.8654 \pm 0.0156 $ & $ 0.8175 \pm 0.0889 $ & $ 0.5022 \pm 0.0602 $ & $ 0.9767 \pm 0.0105 $ & $ 0.5911 \pm 0.0556 $ \\
& (0.1) - (0.6) & $ 0.8661 \pm 0.0146 $ & $ 0.7953 \pm 0.0742 $ & $ 0.5033 \pm 0.0617 $ & $ 0.9773 \pm 0.0097 $ & $ 0.5854 \pm 0.0544 $ \\
& (0.1) - (0.7) & $ 0.8651 \pm 0.0179 $ & $ 0.8051 \pm 0.0975 $ & $ 0.4822 \pm 0.0643 $ & $ \bm{0.9827 \pm 0.0087} $ & $ 0.5744 \pm 0.0642 $ \\
\hline

\multirow{ 9}{*}{\rotatebox{90}{DFML}}
& (0.01) - (0.5) & \cellcolor{black!25}$ \bm{0.8984 \pm 0.0153} $ & $ 0.8517 \pm 0.0566 $ & $ \bm{0.7033 \pm 0.0646} $ & $ 0.9581 \pm 0.0193 $ & $ \bm{0.7350 \pm 0.0497} $ \\
& (0.01) - (0.6) & $ 0.8932 \pm 0.0149 $ & $ 0.8526 \pm 0.0614 $ & $ 0.6944 \pm 0.0517 $ & $ 0.9544 \pm 0.0159 $ & $ 0.7280 \pm 0.0474 $ \\
& (0.01) - (0.7) & $ 0.8864 \pm 0.0101 $ & $ 0.8381 \pm 0.0747 $ & $ 0.6189 \pm 0.0372 $ & $ 0.9683 \pm 0.0127 $ & $ 0.6801 \pm 0.0361 $ \\
& (0.05) - (0.5) & $ 0.8933 \pm 0.0176 $ & $ 0.8429 \pm 0.0782 $ & $ 0.7022 \pm 0.0655 $ & $ 0.9514 \pm 0.0199 $ & $ 0.7295 \pm 0.0538 $ \\
& (0.05) - (0.6) & $ 0.8934 \pm 0.0175 $ & $ 0.8459 \pm 0.0592 $ & $ 0.6856 \pm 0.0670 $ & $ 0.9570 \pm 0.0165 $ & $ 0.7238 \pm 0.0558 $ \\
& (0.05) - (0.7) & $ 0.8949 \pm 0.0148 $ & \cellcolor{black!25}$ \bm{0.8696 \pm 0.0661} $ & $ 0.6567 \pm 0.0572 $ & $ 0.9679 \pm 0.0102 $ & $ 0.7195 \pm 0.0461 $ \\
& (0.1) - (0.5) & $ 0.8945 \pm 0.0139 $ & $ 0.8431 \pm 0.0848 $ & $ 0.6722 \pm 0.0558 $ & $ 0.9627 \pm 0.0134 $ & $ 0.7148 \pm 0.0492 $ \\
& (0.1) - (0.6) & $ 0.8900 \pm 0.0172 $ & $ 0.8279 \pm 0.0588 $ & $ 0.6733 \pm 0.0646 $ & $ 0.9565 \pm 0.0168 $ & $ 0.7117 \pm 0.0532 $ \\
& (0.1) - (0.7) & $ 0.8946 \pm 0.0177 $ & $ 0.8691 \pm 0.0840 $ & $ 0.6344 \pm 0.0641 $ & $ \bm{0.9747 \pm 0.0134} $ & $ 0.7009 \pm 0.0574 $ \\
\hline

\multirow{ 9}{*}{\rotatebox{90}{DFML$_{\chi^2}$}}
& (0.01) - (0.5) & $ 0.8887 \pm 0.0241 $ & $ 0.7589 \pm 0.0598 $ & $ 0.8833 \pm 0.0729 $ & $ 0.8901 \pm 0.0262 $ & $ 0.7884 \pm 0.0549 $ \\
& (0.01) - (0.6) & $ 0.8790 \pm 0.0213 $ & $ 0.7390 \pm 0.0530 $ & $ 0.8733 \pm 0.0567 $ & $ 0.8810 \pm 0.0223 $ & $ 0.7708 \pm 0.0483 $ \\
& (0.01) - (0.7) & $ 0.8888 \pm 0.0180 $ & $ 0.7773 \pm 0.0569 $ & $ 0.8044 \pm 0.0543 $ & $ \bm{0.9146 \pm 0.0220} $ & $ 0.7568 \pm 0.0391 $ \\
& (0.05) - (0.5) & $ 0.8806 \pm 0.0185 $ & $ 0.7317 \pm 0.0585 $ & \cellcolor{black!25}$ \bm{0.8978 \pm 0.0602} $ & $ 0.8751 \pm 0.0186 $ & $ 0.7819 \pm 0.0510 $ \\
& (0.05) - (0.6) & $ 0.8859 \pm 0.0181 $ & $ 0.7567 \pm 0.0408 $ & $ 0.8900 \pm 0.0700 $ & $ 0.8845 \pm 0.0243 $ & \cellcolor{black!25}$ \bm{0.7885 \pm 0.0425} $ \\
& (0.05) - (0.7) & $ 0.8918 \pm 0.0194 $ & $ \bm{0.7792 \pm 0.0435} $ & $ 0.8389 \pm 0.0615 $ & $ 0.9079 \pm 0.0176 $ & $ 0.7808 \pm 0.0476 $ \\
& (0.1) - (0.5) & $ 0.8856 \pm 0.0223 $ & $ 0.7485 \pm 0.0534 $ & $ 0.8911 \pm 0.0590 $ & $ 0.8837 \pm 0.0284 $ & $ 0.7879 \pm 0.0492 $ \\
& (0.1) - (0.6) & $ 0.8780 \pm 0.0221 $ & $ 0.7425 \pm 0.0463 $ & $ 0.8722 \pm 0.0495 $ & $ 0.8799 \pm 0.0259 $ & $ 0.7726 \pm 0.0420 $ \\
& (0.1) - (0.7) & $ \bm{0.8924 \pm 0.0159} $ & $ 0.7756 \pm 0.0528 $ & $ 0.8422 \pm 0.0596 $ & $ 0.9078 \pm 0.0186 $ & $ 0.7812 \pm 0.0432 $ \\
\hline

\multirow{ 9}{*}{\rotatebox{90}{DFML$_{Lap}$}}
& (0.01) - (0.5) & $ \bm{0.8687 \pm 0.0194} $ & $ 0.8075 \pm 0.0885 $ & $ \bm{0.5256 \pm 0.0729} $ & $ 0.9737 \pm 0.0124 $ & $ \bm{0.6060 \pm 0.0706} $ \\
& (0.01) - (0.6) & $ 0.8643 \pm 0.0165 $ & $ 0.7908 \pm 0.1072 $ & $ 0.5100 \pm 0.0639 $ & $ 0.9727 \pm 0.0095 $ & $ 0.5885 \pm 0.0722 $ \\
& (0.01) - (0.7) & $ 0.8605 \pm 0.0139 $ & $ 0.7845 \pm 0.1009 $ & $ 0.4811 \pm 0.0588 $ & $ 0.9768 \pm 0.0117 $ & $ 0.5686 \pm 0.0582 $ \\
& (0.05) - (0.5) & $ 0.8613 \pm 0.0190 $ & $ 0.8091 \pm 0.0913 $ & $ 0.4956 \pm 0.0692 $ & $ 0.9733 \pm 0.0106 $ & $ 0.5848 \pm 0.0670 $ \\
& (0.05) - (0.6) & $ 0.8607 \pm 0.0170 $ & $ 0.8033 \pm 0.0887 $ & $ 0.4900 \pm 0.0585 $ & $ 0.9744 \pm 0.0112 $ & $ 0.5785 \pm 0.0631 $ \\
& (0.05) - (0.7) & $ 0.8682 \pm 0.0181 $ & $ \bm{0.8214 \pm 0.0830} $ & $ 0.5089 \pm 0.0649 $ & $ 0.9783 \pm 0.0100 $ & $ 0.6036 \pm 0.0618 $ \\
& (0.1) - (0.5) & $ 0.8610 \pm 0.0160 $ & $ 0.8028 \pm 0.0827 $ & $ 0.4822 \pm 0.0648 $ & $ 0.9771 \pm 0.0108 $ & $ 0.5724 \pm 0.0570 $ \\
& (0.1) - (0.6) & $ 0.8621 \pm 0.0144 $ & $ 0.7912 \pm 0.0791 $ & $ 0.4844 \pm 0.0607 $ & $ 0.9779 \pm 0.0091 $ & $ 0.5702 \pm 0.0565 $ \\
& (0.1) - (0.7) & $ 0.8604 \pm 0.0157 $ & $ 0.7901 \pm 0.0913 $ & $ 0.4600 \pm 0.0540 $ & \cellcolor{black!25}$ \bm{0.9834 \pm 0.0089} $ & $ 0.5541 \pm 0.0554 $ \\
\hline

\hline
\end{tabular}
}
\end{table*}


\section*{References}
\bibliographystyle{elsarticle-harv}
\bibliography{references}





\end{document}